\documentclass[natbib]{svjour3}  
\smartqed 

\usepackage{natbib}
\usepackage{graphicx}
\usepackage[ruled,linesnumbered,lined,noend]{algorithm2e}
\usepackage{lineno}
\usepackage{graphicx}
\usepackage{subfigure}
\usepackage{pifont} 
\usepackage{url}
\usepackage{color}
\usepackage{booktabs}
\usepackage{multirow}
\usepackage{latexsym,bm,amsmath,amssymb}
\usepackage{array}
\usepackage{algpseudocode}
\usepackage{lscape}
\usepackage{balance}
\usepackage[colorlinks,linkcolor = blue,citecolor = blue,urlcolor=blue]{hyperref}
\usepackage{tcolorbox}
\usepackage{ragged2e}
\usepackage{mathptmx}      

\newcommand{\yg}[1]{\textcolor{black}{#1}}

\definecolor{light-gray}{gray}{0.95}

\newcommand{\code}[1]{\colorbox{light-gray}{\texttt{#1}}}
\newcommand{\tool}{\textsf{TurduckenGen}} 
\definecolor{light-gray}{gray}{0.85}
\newcommand{\tabincell}[2]{\begin{tabular}{@{}#1@{}}#2\end{tabular}}  
\definecolor{mygray}{gray}{.9}
\journalname{Empirical Software Engineering}
\begin{document}

\title{A Syntax-Guided Multi-Task Learning Approach for Turducken-Style Code Generation}

\titlerunning{Turducken-Style Code Generation}        

\author{
	Guang Yang \and Yu Zhou \and Xiang Chen \and Xiangyu Zhang \and Yiran Xu 
	\and Tingting Han \and Taolue Chen
}

\authorrunning{Yang et al} 

\institute{
Guang Yang \at 
The College of Computer Science and Technology,\\
Nanjing University of Aeronautics and Astronautics, Nanjing, China\\
\email{novelyg@outlook.com}
\and
Yu Zhou \at
The College of Computer Science and Technology,\\
Nanjing University of Aeronautics and Astronautics, Nanjing, China\\
\email{zhouyu@nuaa.edu.cn}
\and
Xiang Chen \at
The School of Information Science and Technology\\
Nantong University, Nantong, China\\
\email{xchencs@ntu.edu.cn}
\and
Xiangyu Zhang \at
The College of Computer Science and Technology,\\
Nanjing University of Aeronautics and Astronautics, Nanjing, China\\
\email{zhangx1angyu@nuaa.edu.cn}
\and
Yiran Xu \at
The College of Computer Science and Technology,\\
Nanjing University of Aeronautics and Astronautics, Nanjing, China\\
\email{behumble@nuaa.edu.cn}
\and
Tingting Han \at
Department of Computer Science,\\
Birkbeck, University of London, UK\\
\email{t.han@bbk.ac.uk}
\and
Taolue Chen \at
Department of Computer Science,\\
Birkbeck, University of London, UK\\
\email{t.chen@bbk.ac.uk}
}

\date{Received: date / Accepted: date}
\maketitle

\begin{abstract}
Due to the development of pre-trained language models, automated code generation techniques have shown great promise in recent years.
However, the generated code is difficult to meet the syntactic constraints of the target language,
especially in the case of Turducken-style code, where declarative code snippets are embedded within imperative programs. 
In this study, we summarize the lack of syntactic constraints into three significant challenges: (1) the efficient representation of syntactic constraints, (2) the effective integration of syntactic information, and (3) the scalable syntax-first decoding algorithm.
To address these challenges, we propose a syntax-guided multi-task learning approach {\tool}.
Specifically, we first explicitly append the type information to the code tokens to capture the representation of syntactic constraints.
Then we formalize code generation with syntactic constraint representation as an auxiliary task to enable the model to learn the syntactic constraints of the code.
Finally, the syntactically correct code is selected accurately from the multiple candidates with the help of the compiler feedback.
Extensive experiments and comprehensive analysis demonstrate the effectiveness and general applicability of our approach after being compared with six state-of-the-art baselines on two Turducken-style code datasets. 
Finally, we conducted a human study and found the code quality generated by our approach is better than baselines in terms of code readability and semantic similarity.

\end{abstract}

\section{Introduction}
\label{sec:intro}

Contemporary society is dependent on intricate software applications, and the development of such applications is a complex and prolonged process~\citep{liu2023reliability}. As the complexity of software increases, the development process becomes increasingly time-consuming and susceptible to errors. Moreover, the demand for acquiring proficiency in multiple programming languages is high, especially for novice developers, which further complicates software development~\citep{xu2022ide}. To mitigate these challenges, neural code generation has been proposed as a solution that aims to synthesize code snippets based on functional descriptions, with the potential to alleviate the burden of programmers during the development process.

Based on the type of target programming languages, the existing code generation tasks can be classified into imperative code generation and declarative code generation. Previous studies have demonstrated that existing methods can generate code with an accuracy of over 90\% on datasets of declarative programs~\citep{sun2020treegen, xuan2021sead}, while the accuracy on datasets of imperative programs is less than 35\%~\citep{ahmad2021unified, wang2021codet5}. However, large software systems are rarely developed exclusively in a declarative language in practical software development. Instead, declarative programs are commonly embedded within imperative programs, which are normally referred to as \emph{Turducken-style programs}~\citep{liang2021lyra}. For example, in information management systems, developers often include SQL statements within imperative programs; in data mining systems, developers often incorporate regular expressions within imperative programs. A scientifically compelling and engineering-significant question is how to bring the good performance of automatic declarative program generation to real-world software development.

In general, automatic code generation can be beneficial in various aspects, such as increasing development efficiency, reducing potential faults in code, and enhancing code maintainability and readability. The automatic Turducken-style code generation method is primarily intended to assist programmers who need to simultaneously write declarative and imperative programs during the development process, thereby increasing development efficiency and productivity.

\begin{figure}[htbp]
\centering
\vspace{-2mm}
\includegraphics[width=0.95\textwidth]{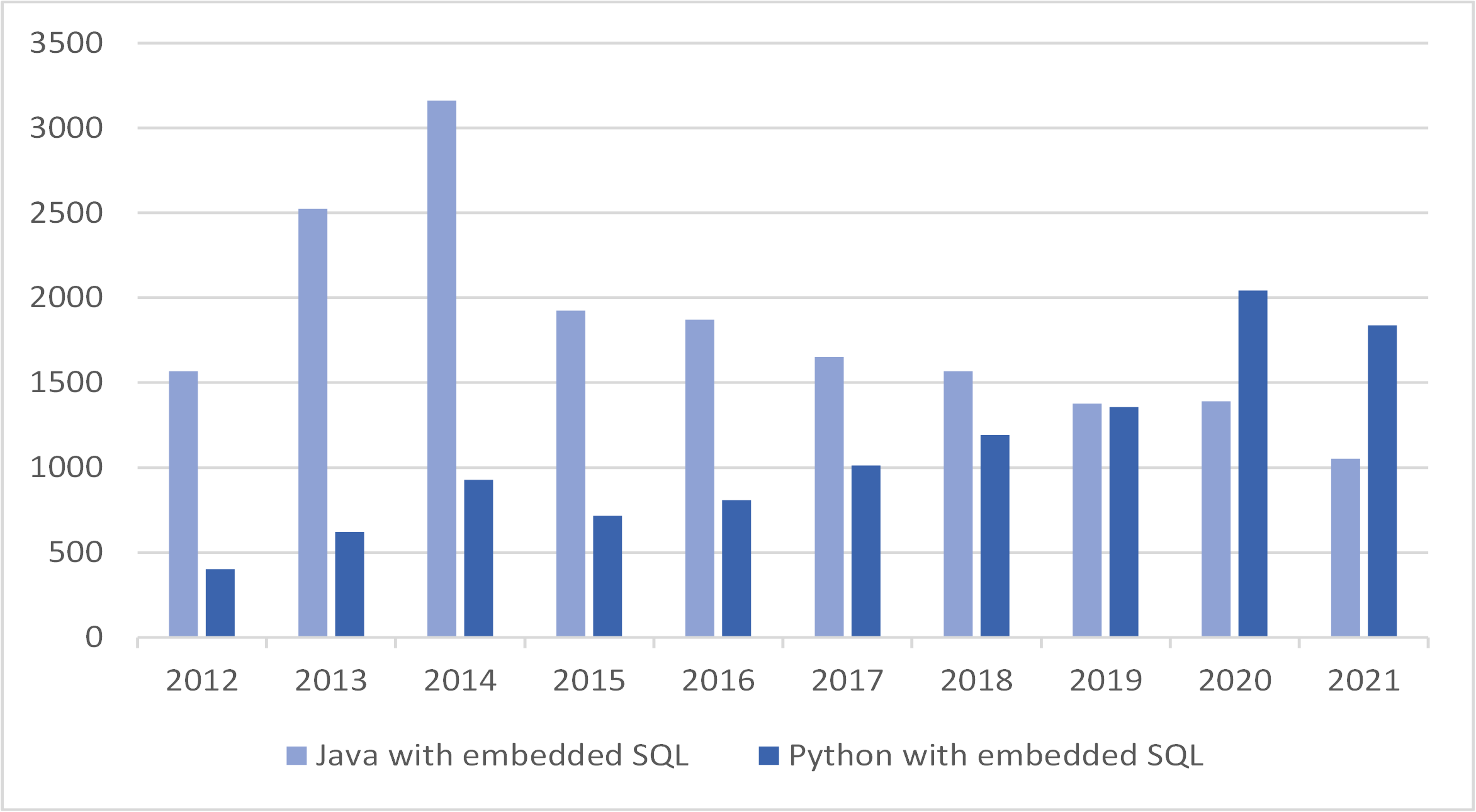}
\caption{The number of related posts in Stack Overflow by year}
\vspace{-2mm}
\label{fig:SO}
\end{figure}

Turducken-style code is prevalent in real-world software development. \citep{allamanis2013and} observed that SQL is frequently used in conjunction with imperative programming languages, such as Java and Python. 
Moreover, we have retrieved Q\&A posts on Stack Overflow with both keywords \textit{SQL and Java} or \textit{SQL and Python} in the past 10 years. The statistics are shown in Fig.~\ref{fig:SO} where we can find that the number of posts tagged by \textit{SQL and Java} has 
been stable, while the number of posts tagged by \textit{SQL and Python} has increased steadily due to the growing popularity of Python.
Table~\ref{tab:examples} presents a real-world post from Stack Overflow\footnote{https://stackoverflow.com/questions/29286725/}
In this post, the user had successfully constructed a SQL statement query, but encountered difficulties in embedding it into a Java program. This illustrates the challenges faced by developers who are not proficient in imperative programming languages when attempting to write Turducken-style code in the software development process.

\begin{table}[htbp]
\caption{A Post Related to Turducken-style Code Generation on Stack Overflow}
\begin{tabular}{ll}
\toprule
\textbf{Title} & Using Spring JdbcTemplate to extract one string\\
\midrule
\textbf{Content} & \begin{tabular}[c]{@{}l@{}}
Can't seem to find a way to get one string from table using JdbcTemplate query.\\
This is the table my sql returns:  \\
\code{ID | STREET\_NAME\quad\quad\quad\quad\quad\quad} \\
\code{----------------------------} \\
\code{1~ | Elm street~\quad\quad\quad\quad\quad\quad} \\ 
Now how am I supposed to get the value of STREET\_NAME. SQL always returns one row, \\ so no need to worry about returning more than one row.  \\
But how can I extract ``Elm street" from it using JdbcTemplate? 
\end{tabular}\\
\midrule
\textbf{Tags} & java, sql, hsqldb, jdbctemplate  \\
 \bottomrule
\end{tabular}
\label{tab:examples}
\end{table}

In previous studies~\citep{ahmad2021unified, wang2021codet5, niu2022spt, chakraborty2022natgen}, a prevalent approach is to directly fine-tune pre-trained language models (PLMs) to generate code. However, this approach has a severe limitation, i.e., the generated code may not follow the syntactic rules of the targeted programming language~\citep{https://doi.org/10.48550/arxiv.2211.00818}, which can result in the failed compilation. This issue is particularly serious for domain-specific datasets. 
For example, in the field of exploit code generation, the two previous studies~\citep{liguori2021evil,yang2023exploitgen} both exhibited that even the most advanced pre-training models can generate code that is lexically similar. Therefore, there is still room for improvement in terms of grammatical accuracy. 
In data science code generation, \citep{huang2022execution} found that about 8\% of the errors in the generated code are due to syntax issues.
Similarly, \citep{liang2021lyra} have demonstrated the feasibility of using pre-trained language models to generate Turducken-style code but also admitted the less accurate syntactic conformation.
The purpose of the automatic code generation model is to improve the efficiency of software development through automation and intelligent techniques. However, if the generated code has syntax problems, developers may need to spend a lot of time fixing the defects in the code.
Therefore, generating code that conforms to syntactic constraints is still a significant area of research within the field.

\smallskip
\noindent\textbf{Proposed solution.} 
To address these issues, we propose a novel approach, i.e., {\tool}, which is based on CodeT5~\citep{wang2021codet5} and multi-task learning. 
{\tool} proposes corresponding solutions from three perspectives.
(1) \textbf{The efficient representation of syntactic constraints.} {\tool} uses the syntax rule description language to represent syntactic constraints. {\tool} first parses the code into an abstract syntax tree, transforming it from a serialized text representation into a syntax tree with rich structural information. Then {\tool} proposes the syntax augmented traversal (SAT) algorithm, which can capture the representation of syntactic constraints by traversing the abstract syntax tree of the original code and explicitly appending the type information of the parent node to the code tokens. We refer to code with syntactically constrained representations as syntax-guided code.
(2) \textbf{The effective integration of syntactic information.} {\tool} formalizes syntax-guided code generation as an auxiliary task to enable the model to learn to generate code that adheres to syntactic constraints. Moreover, {\tool} employs a hard prompt method to facilitate the model's ability to distinguish the primary tasks and the auxiliary tasks. 
{\tool} employs the pre-training model CodeT5 to understand the user's functional description and learn both the target code and its corresponding code satisfying syntactic constraints. The encoder and decoder parameters are shared in the joint learning process, and the task-specific layers are customized in the final stage of mapping the semantic vector to the vocabulary. In order to incorporate the syntactic constraints learned in the auxiliary task into the primary task, {\tool} utilizes a gated linear unit for knowledge fusion. 
(3) \textbf{The scalable syntax-first decoding algorithm.} {\tool} proposes the syntax-first beam search (SF-Beam Search) method to maximize the ability of the model to generate syntax-correct code, where it can be easily integrated into existing pre-trained models. Specifically, the syntactically correct code can be selected accurately from the multiple candidates with the help of the compiler feedback.

To evaluate the effectiveness of {\tool}, we conduct experiments on two Turducken-style code datasets Lyra and Pisces, where Lyra~\citep{liang2021lyra} is obtained from GitHub repositories and annotated by human annotators, whereas Pisces is obtained by manually translating the python code in Lyra into the corresponding Java code through a crowd-sourcing approach.
{\tool} is compared to six state-of-the-art baselines, including Transformer~\citep{vaswani2017attention}, CodeBERT~\citep{feng2020codebert}, GraphCodeBERT~\citep{guo2021graphcodebert}, GPT~\citep{radford2019language}, CodeGPT~\citep{lu2021codexglue}, \yg{UniXcoder}~\citep{guo2022unixcoder} and CodeT5~\citep{wang2021codet5}, in terms of six automatic performance metrics (i.e., BLEU~\citep{papineni2002bleu}, Weighted BLEU~\citep{ren2020codebleu}, Crystal BLEU~\citep{eghbali2022crystalbleu}, \yg{Code BLEU}~\citep{ren2020codebleu}, Syntax Match~\citep{ren2020codebleu}, Syntax Exact Match~\citep{liang2021lyra}, and Code Executable~\citep{liang2021lyra}). The comparison results show that {\tool} outperforms these baselines. 
Moreover, we design ablation studies to verify the effectiveness of multi-task learning and the method SF-Beam Search.
Finally, from practitioners’ perspectives on the generated Turducken-style code, we conduct a human evaluation to evaluate the quality of the generated code in terms of code readability and semantic similarity. The final results also show the competitiveness of our proposed approach.
 
The main contributions of our study can be summarized as follows.
\begin{itemize}

\item  We propose a novel approach {\tool} for generating syntax-guided Turducken-style code. By utilizing CodeT5 and multi-task learning, we can effectively integrate syntax knowledge into the generated code, while our proposed SAT algorithm and SF-Beam Search method can improve the ability of the model to capture the representation of syntactic constraints and generate syntax-correct code.

\item We conduct comprehensive experiments using both automatic evaluation metrics and human evaluation to assess the performance of {\tool} on two Turducken-style code datasets. The results of the evaluation indicate that {\tool} outperforms the state-of-the-art baselines.

\item To facilitate the replication and reuse of {\tool}, we develop an Integrated Development Environment (IDE) plug-in and make our source code, trained models, as well as the datasets in the GitHub repository publicly available\footnote{\url{https://github.com/NTDXYG/TurduckenGen}}.

\end{itemize}

\smallskip
\noindent\textbf{Structure.} 
The rest of the paper is organized as follows. 
Section~\ref{sec:ground} introduces the background of our work.
Section~\ref{sec:method} describes the framework of {\tool} and its key components. 
Section~\ref{sec:setup} presents the experimental design and result analysis. 
Human Evaluation and potential threats to validity are given in Section~\ref{sec:discussion} and Section~\ref{sec:threats} respectively.
Section~\ref{sec:related} reviews the related work and emphasize the novelty of our study, and Section~\ref{sec:conclusion} concludes the paper.

\section{Background} 
\label{sec:ground}

In this section, we provide an overview of the background of Turducken-style code, CodeT5, and multi-task learning.

\subsection{Turducken-style Code}

In the neural code generation task, programming languages are typically classified into two categories: declarative and imperative programming languages. 
Declarative programming is a paradigm that expresses the logic of a computation without describing its control flow~\citep{lloyd1994practical}. It can simplify writing parallel programs~\citep{bailey2009workshop}. Examples of declarative languages include database query languages (e.g., SQL, XQuery), regular expressions, markup languages (e.g., HTML, XAML), functional programming (e.g. Haskell, Scheme), and configuration management systems (e.g., Json, YAML).
On the other hand, imperative programming is a paradigm that focuses on describing how a program should accomplish a task without specifying all the details of how the program should achieve the result~\citep{gifford1986integrating}. Imperative programming focuses on describing how a program runs step-by-step, rather than on a high-level description of its intended outcome. Examples of imperative languages include C, C++, Java, Python, and so on.

Turducken-style code, which was first proposed by \citep{liang2021lyra}, refers to a style of code where declarative programming is embedded within imperative programming. This type of code can be commonly found in real-world software systems. For example, SQL statements are often used with imperative programming in information management systems, and regular expressions are commonly used with imperative programming in data mining systems.

\subsection{CodeT5}

CodeT5~\citep{wang2021codet5} is an identifier-aware unified pre-trained encoder-decoder model for code understanding and generation, which is based on the Transformer architecture \citep{vaswani2017attention}. It is pre-trained on the large-scale dataset CodeSearchNet~\citep{husain2019codesearchnet} and BigQuery~\citep{fernandes2015bigquery}, which includes eight programming languages (i.e., Go, Java, Javascript, PHP, Python, Ruby, C, and CSharp). To address out-of-vocabulary issues, CodeT5 utilizes a code-specific tokenizer based on the Byte-level BPE method. To take advantage of both bimodal and unimodal large-scale data, CodeT5 proposes four pre-training tasks: Masked Span Prediction (MSP), Identifier Tagging (IT), Masked Identifier Prediction (MIP), and Bimodal Dual Generation (BDG).

The MSP pre-training task takes a lexical perspective of the code, it utilizes a whole word span masking objective that randomly masks spans of arbitrary lengths and then predicts these masked spans combined with some sentinel tokens at the decoder. The IT and MIP pre-training tasks take a syntactic perspective of the code. It aims to notify the model with the knowledge of whether a code token is an identifier or not, and MIP masks all identifiers in the program language segment, which is inspired by obfuscation in the field of software engineering. The BDG pre-training task treats code generation and code summarization as dual tasks. To bridge the gap between the programming language and natural language, BDG leverages bimodal data to train the model for bidirectional conversion.

\subsection{Multi-task Learning}

Multi-task learning (MTL) is a machine learning paradigm. MTL aims to leverage useful information shared across multiple related tasks, which can improve the generalization performance on all tasks or enhance the model performance for a specific task by using auxiliary tasks~\citep{liu2020multi}. MTL methods can be divided into hard or soft parameter sharing. In hard parameter sharing, the hidden layer is shared between all tasks, while keeping the output layer for several specific tasks. Intuitively, the more tasks that are learned simultaneously, the more representation of the tasks can be captured by the model, resulting in less risk of overfitting the primary task. In soft parameter sharing, each task has its own hidden layers and output layer. To ensure that the parameters of each task are similar, the distance between the parameters of each task is regularized.

In the Turducken-style code generation task, MTL has a natural advantage due to the limited labeled data. From a data augmentation perspective, MTL aggregates training samples from datasets of multiple tasks, which is especially beneficial for low-resource tasks whose labeled dataset is sometimes too small to sufficiently train a model. In most cases, the augmented training dataset alleviates the risk of overfitting and leads to more robust models~\citep{sanchez2021rethinking}. Furthermore, MTL provides additional performance gains compared to data augmentation approaches, due to the learned shared knowledge.


\subsection{Prompt-based Learning}

Prompt-based learning~\citep{liu2023pre} is a strategy to train large language models (LLMs) so the same model can be used for different downstream tasks without re-training. Traditional strategies for training large language models (such as GPT-3 and BERT) require the model to be pre-trained with unlabeled data and then fine-tuned for specific tasks with labeled data. In contrast, prompt-based learning models can autonomously tune themselves for different tasks by transferring domain knowledge~\citep{wang2022no} introduced through prompts.
 
In general, a prompt is a snippet of natural language text that is added to unlabeled data during the pre-training phase~\citep{gao2021making}. The art of writing useful prompts is called prompt engineering. According to the flexibility of the inserted prompt, prompt tuning techniques can be categorized into two types, i.e., hard prompt~\citep{gu2022ppt} and soft prompt~\citep{li2021prefix}. The hard prompt is a technique that modifies the model input by adding fixed natural language instruction (prompts), and the tokens in the soft prompt are continuous vectors that can be learned during the tuning stage.

\section{Approach}
\label{sec:method}

\begin{figure*}[htbp]
\centering
\includegraphics[width=1.0\textwidth]{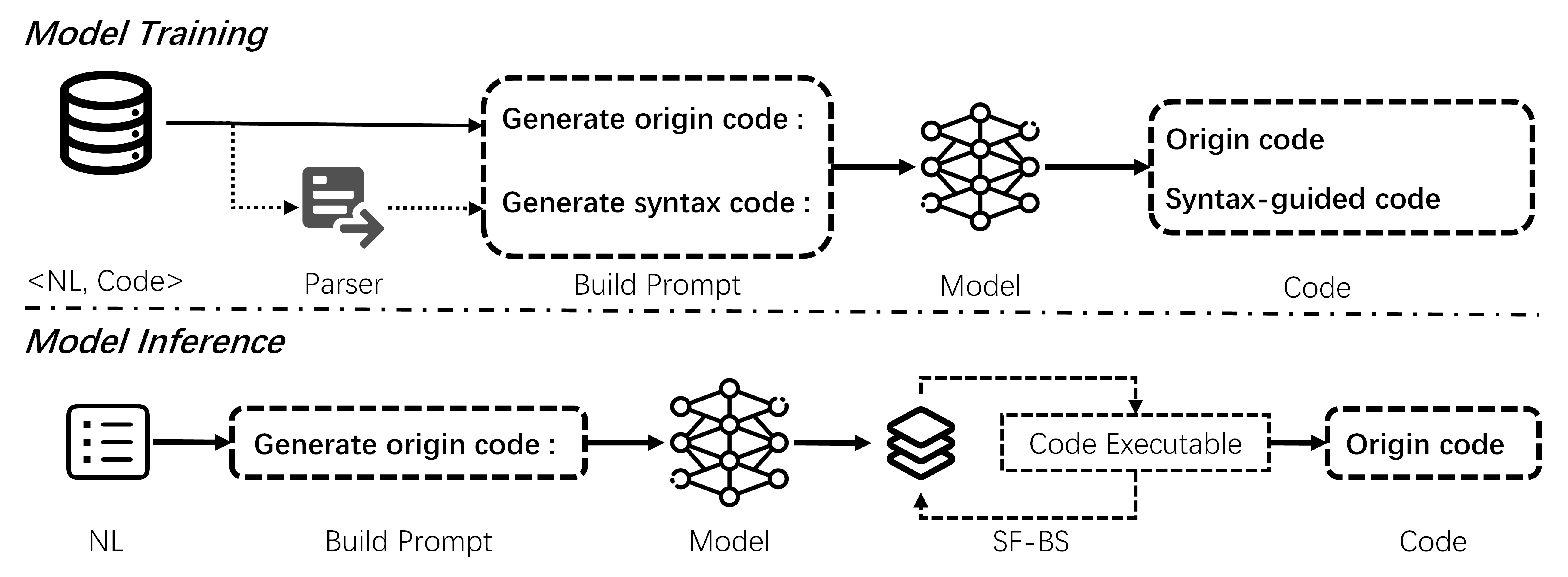}
\caption{Overview framework of our proposed approach}
\label{fig:approach}
\end{figure*}

Fig.~\ref{fig:approach} illustrates the overall framework of {\tool}. It mainly consists of two phases: \textit{Model Training Phase} and \textit{Model Inference Phase}. 
\yg{\textit{Model Training Phase} mainly includes prompt construction, syntax constraint representation, and syntax information integration. Specifically, we build a prompt that consists of the task description. We also represent syntax constraints using a syntax tree and integrate syntax information into the model, which is used to guide the model to generate syntactically correct code.
\textit{Model Inference Phase} mainly introduces the scalable syntax-first decoding algorithm SF-Beam Search. This algorithm is used to generate code by considering the feedback provided by the compiler.} 

\yg{In the rest of this section, we provide more details on these two phases to better illustrate the novelty of our approach.}

\subsection{Model Training Phase}

\yg{To train the model, we input the natural language functional descriptions of both the primary and auxiliary tasks. 
We use the template-based hard prompt method to allow the model to recognize and learn different tasks. 
Compared to continuous-based soft prompt methods, our method is prompted by a series of discrete tokens that are meaningful and understandable.
For our task, we design the template by appending task-specific instructions as follows:}

\begin{equation}\label{prompt}
	\begin{aligned}
		f_{prompt}([TASK], [X], [Y]) &= `` Generate\; [TASK]\; code\; : \; [X] \; [Y]" \\
	\end{aligned}
\end{equation}

\yg{This template explains that the model is to generate the code [Y] according to functional description [X] under the task [TASK]. 
Specifically, the input of the auxiliary task is ``Generate syntax code: [X]", and the output is a sequence that contains syntax information, which is referred to as ``syntax-guided code".
The input of the primary task is ``Generate origin code: [X]", and the output is code sequence, which is referred to as ``original code".
Both the auxiliary task and the primary task use the model's output and ground truth to calculate the loss function, and the model parameters are updated through backward propagation.}

\begin{figure*}[htbp]
	\centering
	\includegraphics[width=1.0\textwidth]{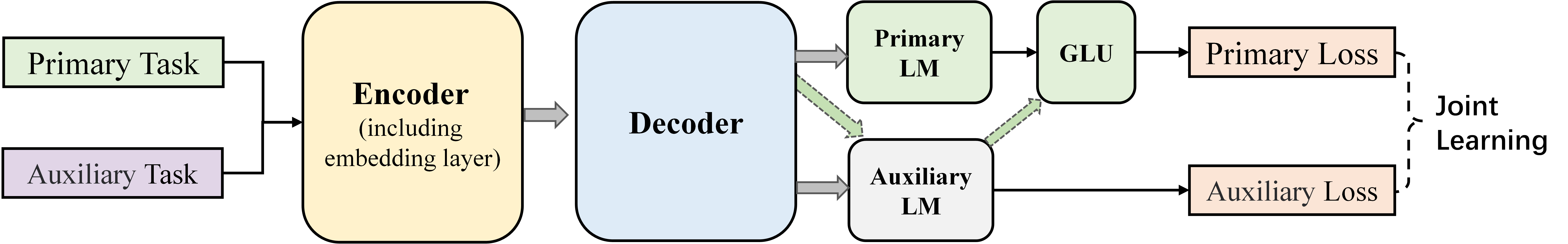}
	\caption{The structure of the multi-task learning used in our neural network}
	\label{fig:structure1}
\end{figure*}

\yg{Fig.~\ref{fig:sat} provides the process of the representation of syntactic constraints, and Fig.~\ref{fig:structure1} provides an overview of the proposed network model architecture for the effective integration of syntactic information. In the rest of this section, we show the details of  the representation of syntactic constraints (i.e., Syntax Augmented Traversal) and the integration of syntactic information (i.e., Model Architecture).}

\subsubsection{Syntax Augmented Traversal}


%



To facilitate the pre-trained language model to generate code that adheres to the target programming language's syntactic constraints, we propose a Syntax Augmented Traversal (SAT) algorithm for the efficient representation of syntactic constraints. This algorithm parses the AST of the source code and then converts it into a sequence of tokens annotated with syntactic information. The details of this algorithm can be found in Algorithm~\ref{alg:sat}.

\textbf{Step 1.} From the root node, we use a pair of XML-like flags to annotate the syntax type.

\textbf{Step 2.} Then, we traverse the sub-trees of the root node by pre-order traversal. 
For a non-leaf node, a determination is made as to whether the node is a string, in which case the label `STR' is utilized for the node. Conversely, the token value to the current node is employed. 
For a leaf node, its root node is incorporated into the XML-like flags.

\textbf{Step 3.} Traversing all sub-trees by recursion, this process continues until all nodes are traversed. The result of this process is a syntax-guided code sequence.

\begin{algorithm}[htbp]\SetKwFunction{SAT}{SAT}
	\small
	\caption{Syntax Augmented Traversal (SAT) Algorithm} \label{alg:sat} 
	\KwIn{
	AST $tree$ of Source Code Parsed by Tree-Sitter;\\}
	\KwOut{
	Syntax-guided Code $syn\_code$;}
        $syn\_code \gets \emptyset$;\\
        \eIf{ ! $tree$.hasChild }{
            \eIf{$tree$.type == Identifier}{
                \eIf{ $tree$.value is String }{
                    $syn\_code \gets$ STR\; 
                } {
                    $syn\_code \gets tree$.value\;
                }
            }{
            $syn\_code \gets tree$.value\;
            }
        } {
            $syn\_code \gets \textless tree$.type$\textgreater$\;
            \For{each $c \in tree.getChilds$} {
                $syn\_code \gets syn\_code$ + \SAT(c)
            }
            $syn\_code \gets \textless$/$tree$.type$\textgreater$\;
        }
    \Return $syn\_code$\; 
\end{algorithm}

\begin{figure*}[htbp]
	\centering
	\includegraphics[width=0.9\textwidth]{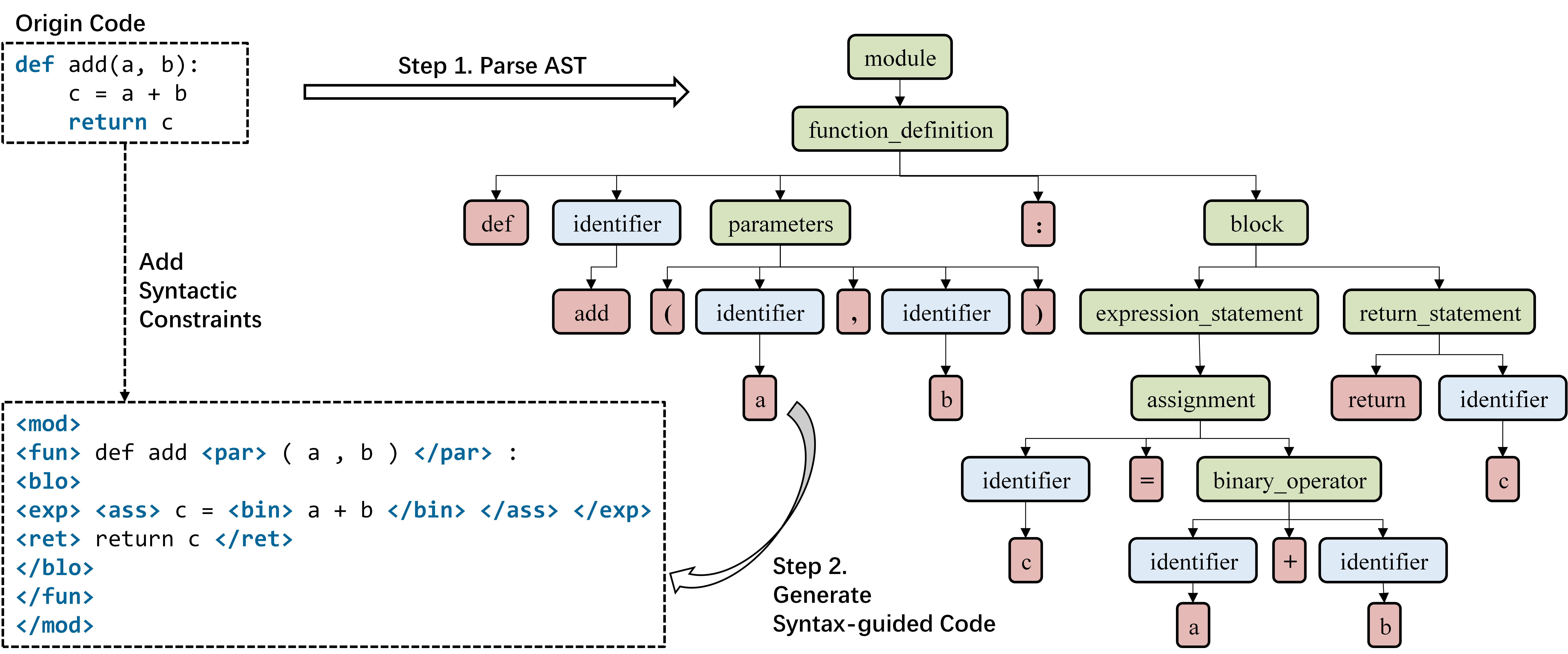}
	\caption{An example is used to illustrate how SAT traverses the AST of the source code}
	\label{fig:sat}
\end{figure*}

An illustrative example is shown in Fig.~\ref{fig:sat}. 
Different colors are assigned to the nodes in the AST in the figure for illustrative purposes. 
Specifically,
the red nodes indicate identifiers in the code, which embody the lexical information of the code. For this type of nodes, their value is retained. 
The blue nodes imply the type of these identifiers. For this type of nodes, their type information is ignored. 
The green nodes represent the complete syntax constraint information of the code (e.g., module, block, and return\_statement), and the first three characters of its type information are extracted as its syntactic information. 
SAT appends these green nodes' type information to the red nodes' tokens explicitly to incorporate syntactic constraints. 
This approach enables the preservation of syntactic information and the ability to revert back to the original code.

%


\subsubsection{Model Architecture}

The architecture of {\tool} adheres to the Transformer~\citep{vaswani2017attention} and pre-trained model CodeT5~\citep{wang2021codet5}, which have been successfully utilized in code-related tasks. The architecture is primarily composed of three sub-modules.
(1) \textbf{Encoder.} It aims to encode the functional description and employs multi-head self-attention to learn the sequential information of the functional description.
(2) \textbf{Decoder.} It aims to generate Turducken-style code through the use of the self-attention layer and the encoder-decoder attention layer.
(3) \textbf{Task-specific Layers.} They are designed to learn the mapping relationships of various tasks separately and to integrate the knowledge of the auxiliary task into the primary task using the Gated Linear Unit (GLU) network.

\noindent \textbf{Encoder.}
{\tool} first generates embedding vectors that capture the semantic meaning of tokens and their position within a functional description. For a functional description $X$, {\tool} firstly tokenizes it into a sequence of sub-words (i.e., $X = {x}_1, \cdots, {x}_m$) by the BPE algorithm~\citep{raffel2020exploring}, where $m$ is the length of the tokenized sequence. 
For each sub-word, {\tool} generates an [1x768] embedding vector and combines it into a matrix to represent the meaningful relationship between a given token and the other tokens.
In addition, to capture the position of each token within a functional description, {\tool} employs the relative position encoding technique~\citep{raffel2020exploring}.

Encoder contains a stack of twelve layers of blocks, each block consists of two sub-components: a multi-head self-attention layer with relative position encoding and a feed-forward neural network. 
In contrast to the original Transformer, each sub-component in {\tool} is followed by a layer normalization and has a residual connection behind it. 
Given an input vector $\boldsymbol{X}$, the first step is to create three main vectors (i.e., a query vector $\boldsymbol{Q}$, a key vector $\boldsymbol{K}$, and a value vector $\boldsymbol{V}$). Further, {\tool} computes the relative positional information $\boldsymbol{P}$, where $\boldsymbol{P}$ is an edge representation for the two inputs in dot-product operation to determine the positional information between tokens. $\boldsymbol{P}$ is supplied to the model as an additional component to the $\boldsymbol{K}$ and $\boldsymbol{V}$, and the final attention score is computed as follows:

\begin{equation}
\operatorname{Attention}(Q, K, V)=\operatorname{softmax}\left(\frac{Q(K+P)^T}{\sqrt{d_k}}\right)(V+P)
\end{equation}

Thus, for the embedding vector $\boldsymbol{X}$, {\tool} first uses a layer normalization step and creates three vectors ($\boldsymbol{Q}$, $\boldsymbol{K}$, and $\boldsymbol{V}$), which are then fed into Multi-head Attention Layer.

\begin{equation}
\operatorname{Q}, \operatorname{K}, \operatorname{V}=\operatorname{LayerNorm}(X)
\end{equation}

Multi-head attention mechanisms obtain $h$ different representations of ($\boldsymbol{Q}$, $\boldsymbol{K}$, and $\boldsymbol{V}$). Then they concatenate the results and project the concatenation with a residual connection layer.

\begin{equation}
\operatorname{head}_i=\text { Attention }\left(Q W_i^Q, K W_i^K, V W_i^V\right)
\end{equation}

\begin{equation}
\operatorname{MultiHead}(Q, K, V)=\text { Concat }_i\left(\operatorname{head}_i\right) W
\end{equation}

\begin{equation}
\operatorname{X}_{\text{atten}} = X + \operatorname{MultiHead}(Q, K, V)
\end{equation}

Finally, {\tool} feeds $\operatorname{X}_{\text{atten}}$ into the feed-forward layer (FFN) to generate the hidden state vector $\operatorname{X}_{\text{hidden}}$.

\begin{equation}
\operatorname{X}_{\text{hidden}} = \operatorname{LayerNorm}(\operatorname{X}_{\text{atten}} + \operatorname{FFN}(X))
\end{equation}

\noindent \textbf{Decoder.} 
In the decoder part, {\tool} also contains a stack of twelve layers of blocks, but each block consists of three sub-components: a masked multi-head self-attention layer with relative position encoding, a multi-head encoder-decoder attention layer with relative position encoding, and a feed-forward neural network. 
The feed-forward neural network is as same as that in the Encoder component, while the self-attention layer is similar to that in the encoder component except that it only deals with generated code tokens in the output sequence. 
Different from the self-attention layer, the encoder-decoder attention layer learns the relationship between the source functional description and the target code.
The calculation of $\operatorname{Y}_{\text{cross-atten}}$ is similar to self-attention. The Queries matrix $\boldsymbol{Q}$ comes from the output of the self-attention layer and the Key matrix $\boldsymbol{K}$ and Values matrix $\boldsymbol{V}$ from the output of the encoder component $\operatorname{X}_{\text{hidden}}$. 
Finally, TurduckenGen feeds $\operatorname{Y}_{\text{cross-atten}}$ into the FFN to generate the hidden state vector $\operatorname{Y}_{\text{hidden}}$.


\medskip
\noindent\textbf{Task-specific Layers.} 
Task-specific output layers are employed to generate task-specific outputs, which contain Auxiliary LM layer, Primary LM layer, and GLU layer. 

For the auxiliary task, we append the prompt `Generate syntax code:' to the functional description, denoted as $X_{\text{aux}}$. Its corresponding decoder output vector is denoted as $\operatorname{Y}_{\text{aux-hidden}}$. 
The auxiliary LM layer is defined to produce the probability distribution of the syntax-guided code.
\begin{equation}
P_{aux} = Y_{\text{aux-hidden}} W^{aux}+b^{syn}
\end{equation}

For the primary task, we append the prompt `Generate origin code:' to the functional description, denoted as $X_{\text{pri}}$. Its corresponding decoder output vector is denoted as $\operatorname{Y}_{\text{pri-hidden}}$. 
Both the Primary LM layer and the Gated Linear Unit (GLU) network are defined to generate the probability distribution of the Turducken-style code.

 \begin{equation}
P_{pri}= \left(Y_{\text{pri-hidden}} W^{pri}+b^{pri}\right) \otimes
\sigma \left(Y_{\text{pri-hidden}} W^{aux}+b^{aux}\right)
\end{equation}

The GLU employed here is intended to integrate the code syntax knowledge obtained from the Auxiliary LM layer into the Primary LM layer for the purpose of enforcing syntactic constraints. 
The GLU allows the network to selectively attend to the output of the semantic vectors in the Auxiliary LM and Primary LM layers for the purpose of knowledge interaction and selection. 
Furthermore, the GLU possesses non-linear characteristics while also maintaining a linear path for the gradient, thereby reducing the issue of vanishing gradients~\citep{dauphin2017language}.

To learn these two tasks jointly, the parameters of {\tool} are trained to minimize the sum of the cross-entropy losses of the two tasks. The final loss function is presented as follows.

\begin{equation}
loss = \min _\theta \mathcal{L}_{Primary}(\theta)+\mathcal{L}_{Auxiliary}(\theta)
\end{equation}


\subsection{Model Inference Phase}

During the model inference phase, the prompt template proposed in Eq.\ref{prompt} is constructed to direct the trained model to generate the original code. 

The model output is the probability of each token. In the decoding phase, we propose a scalable syntax-first beam search (SF-Beam Search) method to generate the Turducken-style code.

Recall that the beam search algorithm~\citep{wiseman2016sequence} maintains a beam of $k$ possible tokens $\left({token}_{t}^i\right){i=1}^k$ at each time step $t$, where $k$ represents the beam size. The possible tokens are updated as follows: 
for each token ${token}_{t}^i$, it adds each of the corresponding $k$ most probable candidates, resulting in at most $k$ × $k$ new tokens of size increased by 1. Then among these tokens, the $k$ tokens with the highest likelihood are selected, thus obtaining the next step tokens $\left({token}_{t+1}^i\right)_{i=1}^k$.
However, the beam search algorithm aims at optimizing likelihood and ignores the code syntax, thus we cannot guarantee that the code decoded by beam search can be compiled and executed correctly.

Our proposed Syntax First Beam Search (SF-Beam Search) method is a simple yet effective solution that can be easily integrated into existing models. The SF-Beam Search method is based on our observation that there is a subtle gap between likelihood and code syntax. The likelihood of a code sequence indicates the maximum probability between tokens, but it does not ensure that the code is syntactically correct.
To fill this gap, the SF-Beam Search method takes the $k$ candidate code generated by the beam search algorithm and feeds them into a compiler tool in descending order of likelihood. If the current candidate code is executable, the SF-Beam Search method outputs it and terminates the loop. If all $k$ candidate code are not executable, the SF-Beam Search method outputs the candidate code with the highest likelihood.

\section{Experimental Setup}
\label{sec:setup}

\subsection{Research Questions}
To evaluate the effectiveness of our proposed approach TurduckenGen, we want to investigate the following four research questions (RQs).

\begin{itemize}
\item \textbf{RQ1}: \textit{Can our proposed approach {\tool} outperform the state-of-the-art baselines?}
\item \textbf{RQ2}: \textit{What is the contribution of the multi-task learning of {\tool}?}
\item \textbf{RQ3}: \textit{What is the benefit of using SF-Beam Search method of {\tool}?}
\item \textbf{RQ4}: \textit{What is the impact of different hard prompts of {\tool}?}
\end{itemize}

In RQ1, we want to compare {\tool} with previous pre-trained code generation methods in terms of 
automatic evaluation metrics.
In RQ2, we want to adopt two variants of the multi-task learning methods to investigate the impacts of the multi-task learning method on {\tool}.
In RQ3, we want to analyze the effectiveness between the SF-Beam Search method in {\tool} and other different decoding methods to demonstrate its benefit. 
In RQ4, we want to further explore the impact of different hard prompts on the performance of {\tool}.

\subsection{Datasets}

The Lyra dataset~\citep{liang2021lyra} which is 
for mapping functional descriptions to Turducken-style code, 
only considers Python code with embedded SQL. 
To improve the diversity of Turducken-style code generation, we used a crowd-sourcing approach to translate the Python code in Lyra into its corresponding Java code and modified 
the functional description. 
Finally, we construct a new high-quality dataset Pisces. The corpus contains 2,000 Turducken-style code and corresponding functional descriptions, which are in Java with embedded SQL.

\begin{figure*}[htbp]
\centering
\includegraphics[width=0.9\textwidth]{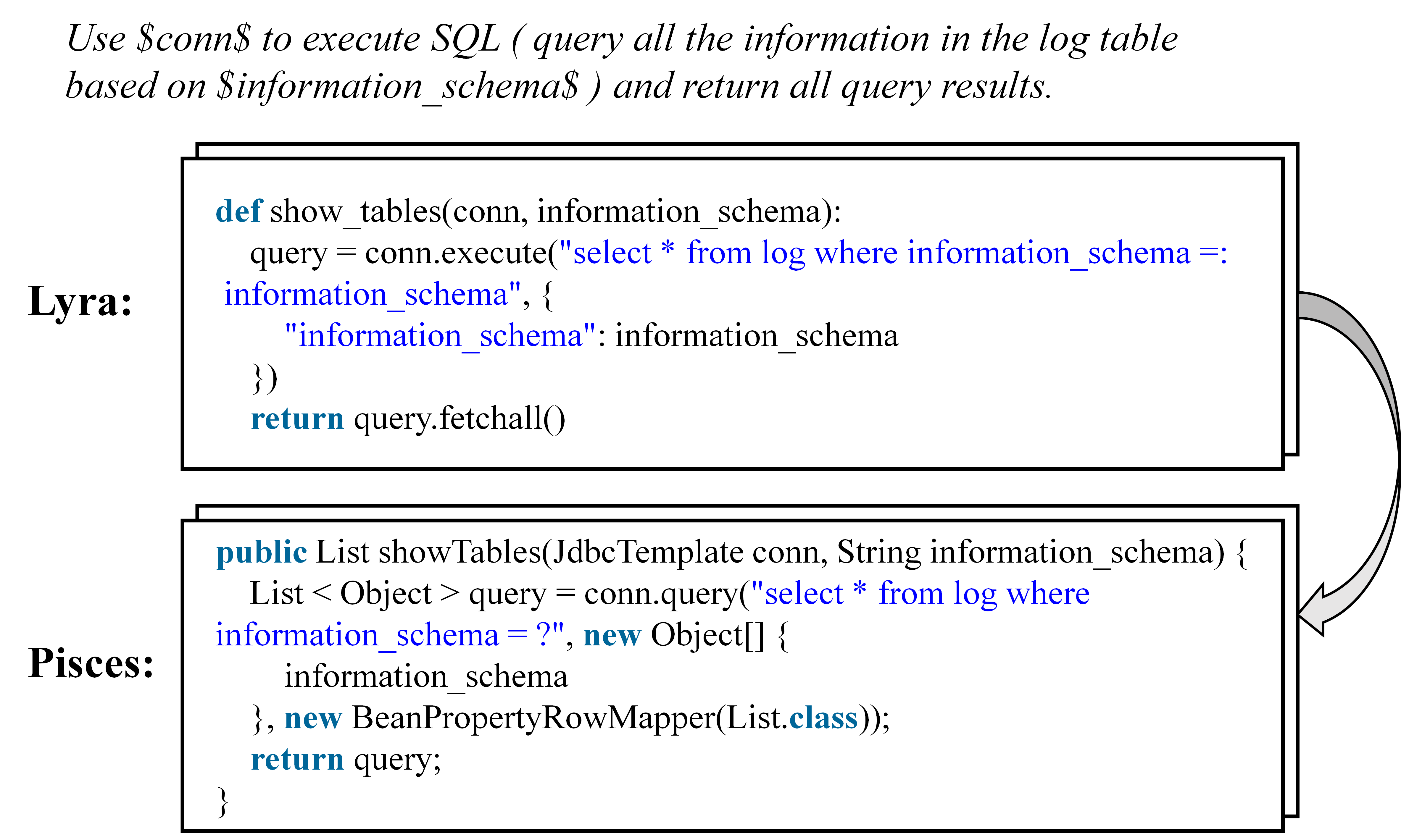}
\caption{An example in the Lyra and Pisces datasets}
\label{fig:dataset}
\end{figure*}

In particular, the dataset Lyra is crawled from GitHub and rewritten code and comments by manual filtering. 
For the dataset Pisces, 
\yg{we hired two Java programmers with 3-5 years of development experience to label the dataset for 10 working days, with each of them spending 1-2 hours per day. They were paid a total of 1,200 Chinese yuan (CNY) as a reward. After the label work was completed, we used Maven to compile and check all the code to ensure the quality of the dataset. }
To ensure the quality of Lyra and Pisces, we also hired other six graduate student volunteers to review this dataset and discuss and fix the disputed data.
Both Lyra and Pisces datasets have two different code styles (i.e., using native SQL and using Object Relational Mapper SQL). To ensure data consistency, ORM in Python uses SQLAlchemy\footnote{\url{https://www.sqlalchemy.org}}, and ORM in Java uses JPA\footnote{\url{https://spring.io/projects/spring-data-jpa}}. 
Fig.~\ref{fig:dataset} shows an example to illustrate our used datasets.

The statistical information (such as the count of code snippets, average tokens in the code, and average tokens in the functional descriptions) of the  Lyra and Pisces datasets is shown in Table~\ref{tab:statistics}.

\begin{table}[htbp]
 \caption{Statistical information of our used Lyra and Pisces datasets}
 \begin{center}
\begin{tabular}{clccc}
\toprule
\textbf{Corpus} & \textbf{Type} & \textbf{Train} & \textbf{Valid} & \textbf{Test} \\ 
\midrule
\multirow{4}{*}{Lyra} & Count & 1,600  & 200 & 200 \\
& Avg. token in NL    &  47.18    &  47.42   &    47.27  \\
& Avg. token in CODE &   57.94    &   58.51   &   57.66    \\
\midrule
\midrule
\multirow{4}{*}{Pisces} & Count & 1,600  & 200 & 200 \\
& Avg. token in NL    &   46.73   &   45.66    &  46.57  \\
& Avg. token in CODE &  79.15  &  89.20   &   84.93  \\
  \bottomrule
\end{tabular}
 \label{tab:statistics}
 \end{center}
\end{table}

\subsection{Evaluation Metrics}

To give a comprehensive evaluation, we use six metrics (i.e., $\mathit{BLEU}$, $\mathit{Weight}$-$\mathit{BLEU}$, $\mathit{Crystal}$-$\mathit{BLEU}$, $\mathit{Syntax}$-$\mathit{Match}$, $\mathit{SyntaxExact}$-$\mathit{Match}$, \yg{$\mathit{Code}$-$\mathit{BLEU}$}, and $\mathit{Code}$-$\mathit{Executable}$), which have been widely used in previous related studies~\citep{hussain2020codegru, hussain2020deep, hussain2021improving, yang2021fine, yang2022dualsc}, to automatically assess the quality of the generated code. 

$\mathit{BLEU}$~\citep{papineni2002bleu}, $\mathit{Weight}$-$\mathit{BLEU}$~\citep{ren2020codebleu}, and $\mathit{Crystal}$-$\mathit{BLEU}$~\citep{eghbali2022crystalbleu} perform code similarity computation from the lexical perspective.
Specifically,
$\mathit{BLEU}$ treats the generated code as the sentence in natural language, $\mathit{Weight}$-$\mathit{BLEU}$ considers keywords in programming languages on the basis of $\mathit{BLEU}$, 
and $\mathit{Crystal}$-$\mathit{BLEU}$ considers the inherent differences between source code and natural language, and optimizes the computation of N-grams in $\mathit{BLEU}$.
$\mathit{Syntax}$-$\mathit{Match}$~\citep{ren2020codebleu} and $\mathit{SyntaxExact}$-$\mathit{Match}$~\citep{liang2021lyra} perform code similarity computation from the syntax perspective.  
$\mathit{Syntax}$-$\mathit{Match}$~\citep{liang2021lyra} obtains all the sub-tree of the AST and then calculates the accuracy by comparing the candidate and reference sub-trees, while $\mathit{SyntaxExact}$-$\mathit{Match}$ is the exact match of both AST and SQL statement, which also means the functional correctness of the generated code.
\yg{$\mathit{Code}$-$\mathit{BLEU}$~\citep{ren2020codebleu} is the mixed evaluation metric, which absorbs the strength of BLEU in the $n$-gram match and further injects code syntax via AST and code semantics via data-flow analysis.}

In addition, $\mathit{Code}$-$\mathit{Executable}$~\citep{wang2022compilable, liang2021lyra} is designed to calculate the executable rate of the generated code. 
\yg{Specially, for the Lyra dataset, we use Python 3.9\footnote{\url{https://www.python.org/downloads/release/python-390/}} as the compiler and utilize the API provided by pylint\footnote{\url{https://github.com/PyCQA/pylint}} for automated checking.
For the Pisces dataset, we use JDK 1.8\footnote{\url{https://www.oracle.com/java/technologies/downloads/\#java8}} as compiler and utilize the command \code{mvn compile} provided by Maven\footnote{\url{https://maven.apache.org/}} for automated checking.}

To ensure the fairness of the experiment, 
we use the scripts provided by \citep{ren2020codebleu} and \citep{liang2021lyra} to calculate the value of these metrics. These performance metrics range from 0 to 1, where larger values represent higher similarity.

\subsection{Baselines}

\yg{To our best knowledge, there are no models specifically designed for Turducken-style code generation. There exists only an empirical study~\citep{liang2021lyra} on Turducken-style code generation and we use their selected methods as our baselines.}
We evaluate the competitiveness of our proposed approach against seven state-of-the-art pre-trained code generation baselines. Specifically, we classify these baselines into four groups.
The first group is the original Transformer~\citep{vaswani2017attention}, which trains the model from the scratch.
The second group is Encoder-Only pre-trained models, including CodeBERT~\citep{feng2020codebert} and GraphCodeBERT~\citep{guo2021graphcodebert}. 
The third group is Decoder-Only pre-trained models, including GPT~\citep{radford2019language} and CodeGPT~\citep{lu2021codexglue}.
The last group is Encoder-Decoder pre-trained model, including \yg{UniXcoder~\citep{guo2022unixcoder}}, CodeT5~\citep{wang2021codet5}.

\subsection{Experimental Settings}

In our empirical study, Transformer is implemented by OpenNMT-py~\citep{klein2017opennmt}, other pre-trained models and corresponding tokenizers are loaded from the official repository Huggingface\footnote{\url{https://huggingface.co/models}} and our used framework is Pytorch\footnote{\url{https://pytorch.org/}}.
The hyper-parameters and their values used in our empirical study are summarized in Table~\ref{Hyper-parameters}, where optimizer, learning rate, and beam size are referenced to the parameters of CodeT5~\citep{wang2021codet5}.

\begin{table}[h]
    \centering
    \caption{Hyperparameter settings}
    \begin{tabular}{c|c||c|c}
    \toprule
       Hyperparameter  & Value &  Hyperparameter  & Value\\
     \midrule
      Optimizer & AdamW & Seed & 1234 \\
      Learning Rate  & 5e-5 & Training batch size & 16 \\
      Beam size & 10 & Validation batch size & 16 \\
      Max input length & 150 & Max output length & 256 \\
      \bottomrule
    \end{tabular}
    \label{Hyper-parameters}
\end{table}

We implement {\tool} using PyTorch 1.8 and all the experiments run on a computer with an Intel(R) Xeon(R) Silver 4210 CPU and the Tesla V100 SXM2 GPU with 32 GB memory. The running OS platform is Linux OS. 

\subsection{Tool Implementation}

\begin{figure}[htbp]
	\centering
  \vspace{-1mm}
	\includegraphics[width=0.95\textwidth]{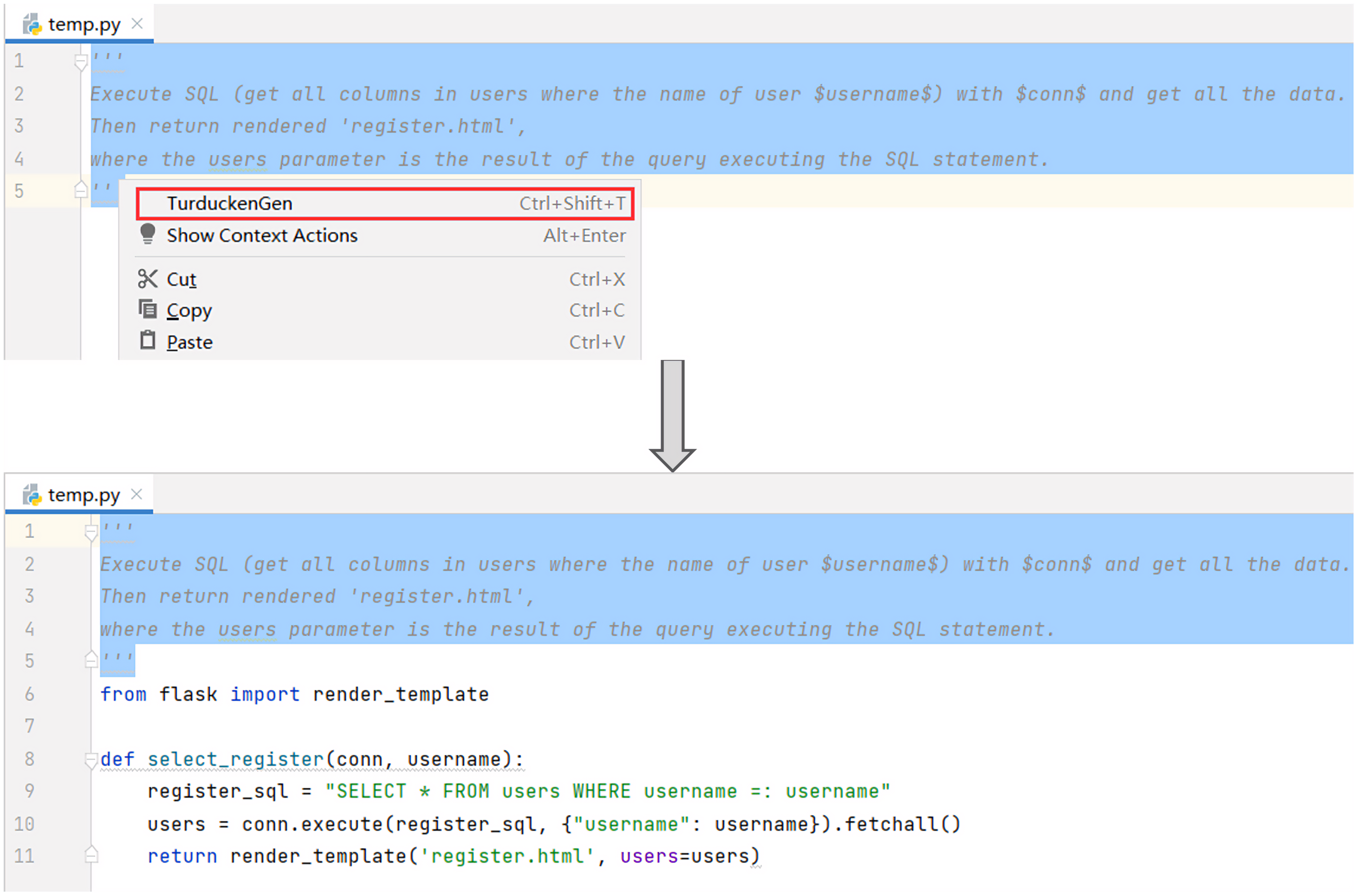}
	\caption{The screenshot of our developed IDE plug-in}
  \vspace{-1mm}
	\label{fig:tool}
\end{figure}

To help developers improve development efficiency, we have also developed an IDE plug-in based on our proposed approach. Our developed plug-in can be integrated into a range of IDEs under JetBrains\footnote{\url{https://www.jetbrains.com/}} products (e.g. IDEA and PyCharm). We show the screenshot of our developed plug-in in Fig.~\ref{fig:tool}. 
In our plug-in, developers first simply enter their functional requirements. Then they select these requirements with the mouse and right-click `TurduckGen' in the shortcut menu. Finally, our plug-in can automatically generate the corresponding code.

\section{Experimental Results}
\label{sec:result}

\subsection{RQ1: Can our proposed approach {\tool} outperform the state-of-the-art baselines?}
\label{sec:resultsRQ1}

We first apply our approach {\tool} and the baseline methods (i.e., Transformer, CodeBERT, GraphCodeBERT, GPT, CodeGPT, and CodeT5) on Lyra and  Pisces, and compare their performance in terms of six automatic evaluation metrics. 
Then we use Wilcoxon signed-rank tests~\citep{wilcoxon1992individual} at the confidence level of 95\% to check whether the performance differences are significant. 
Moreover, we use two samples to perform a qualitative analysis for our proposed approach. 

Table~\ref{tab:RQ1} shows the comparison results between {\tool} and the baselines.
For both the Lyra and Pisces, {\tool} can achieve the best performance in terms of all the performance metrics. 
\yg{Specifically, in terms of $\mathit{BLEU}$, $\mathit{Weight}$-$\mathit{BLEU}$, $\mathit{Crystal}$-$\mathit{BLEU}$, and $\mathit{Code}$-$\mathit{BLEU}$, {\tool} can improve the performance by at least 2.31\%, 2.94\%, 2.84\%, and 3.08\% in Lyra, and by at least 4.46\%, 4.38\%, 10.05\%, and 3.88\% in Pisces.} This demonstrates that  {\tool} can generate more precise code at the lexical level.
In terms of ${Syntax}$-$\mathit{Match}$ and ${SyntaxExact}$-$\mathit{Match}$ metrics, {\tool} can improve the performance by at least 4.47\% and 13.33\% in Lyra, and 3.08\% and 66.67\% in Pisces. This demonstrates that {\tool} is capable of generating code that not only matches syntax more accurately, but also with more similar semantics.
In terms of $\mathit{Code}$-$\mathit{Executable}$ metrics, {\tool} can achieve a 100\% code execution rate on both Lyra and Pisces. This demonstrates that {\tool} generates more syntactically correct code. 

\yg{In addition, we conduct a Wilcoxon signed-rank test~\citep{wilcoxon1992individual} with a significance level of 0.05 to assess the statistical significance of the performance differences between {\tool} and the state-of-the-art baseline CodeT5. 
The Wilcoxon signed-rank test is a two-sided test by default, which means that it tests the null hypothesis that there is no significant difference in performance between {\tool} and CodeT5, regardless of the direction of the difference.
}
In our study, the null hypothesis is denoted as $H_{0}$, i.e., there is no significant difference between {\tool} and the state-of-the-art baseline in terms of metrics $\mathit{BLEU}$, $\mathit{Weight}$-$\mathit{BLEU}$, $\mathit{Crystal}$-$\mathit{BLEU}$, $\mathit{Code}$-$\mathit{BLEU}$, and ${SyntaxMatch}$. 
The results are shown in Table~\ref{tab:RQ1-pvalue}.
In this table, we find all the $p$-values are smaller than $5e\text{-}2$. These statistical results lead to the rejection of the null hypothesis, which means that there exists a significant difference between our approach and baseline.
Note that the results in Table~\ref{tab:RQ1} show that {\tool} can outperform other baselines. Therefore we can conclude that {\tool} achieves better performance than other baseline approaches significantly.

\begin{figure*}[]
\centering
\includegraphics[width=0.75\textwidth]{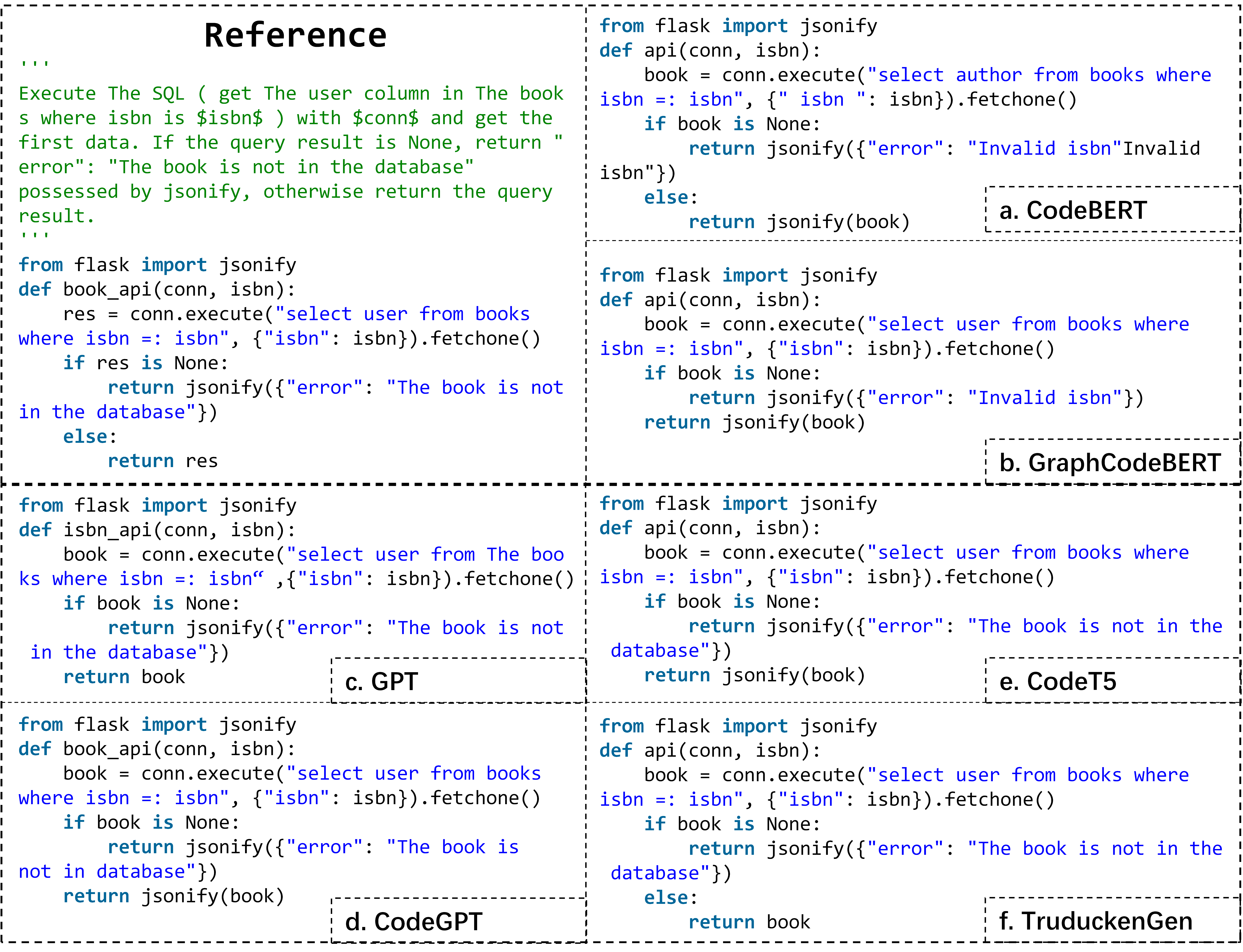}
\caption{The example of Turducken-style code generated by {\tool} and baselines in the Lyra dataset}
\label{fig:Exam-Python}
\end{figure*}

\begin{figure*}[]
\centering
\includegraphics[width=0.75\textwidth]{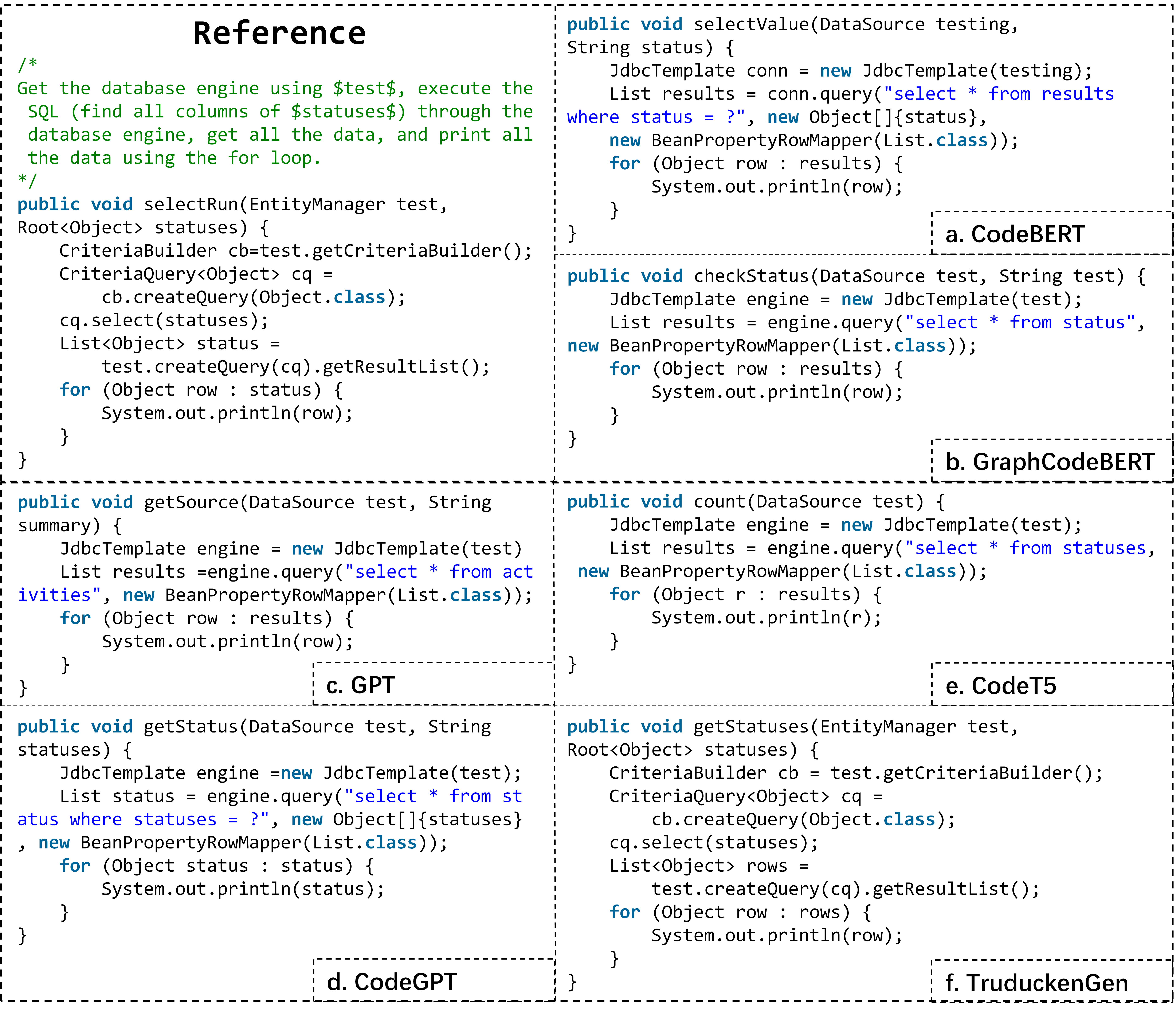}
\caption{The example of Turducken-style code generated  by {\tool} and baselines in the Pisces dataset}
\label{fig:Exam-Java}
\end{figure*}

\begin{table*}[h]
 \caption{The comparison results between our proposed {\tool} and baselines}
     \centering
\scalebox{0.92}{
\begin{tabular}{c|c|ccccccc}
  \toprule
\textbf{Corpus} & 
\textbf{Approach} & \textbf{$\mathit{BLEU}$} & \textbf{$\mathit{Weight}$-$\mathit{B}$} & \textbf{$\mathit{Crystal}$-$\mathit{B}$} & \textbf{$\mathit{Code}$-$\mathit{B}$} & \textbf{$\mathit{S}$-$\mathit{M}$} & \textbf{$\mathit{SE}$-$\mathit{M}$} & \textbf{$\mathit{C}$-$\mathit{E}$}\\
  \midrule
\multirow{7}{*}{Lyra}
& Transformer & 38.97 &39.52 & 10.00 & 46.46 & 48.46 & 0.00 & 15.00  \\
& CodeBERT & 63.29 & 63.93 & 43.70 & 68.77 & 74.04 & 6.50 & 59.50  \\
& GraphCodeBERT & 67.34 & 67.91 & 50.43 & 71.78 & 76.26 & 8.50 & 53.50 \\
& GPT & 72.76 & 73.17 & 60.88 & 77.26 & 81.36 & 21.50 & 92.50 \\
& CodeGPT & 73.23 & 73.65 & 63.51 & 77.41 & 80.37 & 24.00 & 96.00 \\
& UniXcoder & 70.45 & 70.33 & 62.10 & 70.17 & 70.73 & 18.50 & 70.00 \\
& CodeT5 & 76.59 & 76.97 & 67.84 & 80.25 & 83.20 & 30.00 & 95.50 \\
& {\tool}  & \textbf{78.36} & \textbf{79.23} & \textbf{69.77} & \textbf{82.72} &  \textbf{86.92} & \textbf{34.00} & \textbf{100.00} \\
  \midrule
  \midrule
\multirow{7}{*}{Pisces} 
& Transformer & 41.66 & 42.12 & 10.91 & 49.35 & 52.86 & 1.00 & 74.00  \\
& CodeBERT & 57.65 & 58.48 & 32.67 & 61.05 & 61.81 & 0.00 & 93.50  \\
& GraphCodeBERT & 60.22 & 61.05 & 38.06 & 63.16 & 62.68 & 1.00 & 93.50 \\
& GPT & 61.88 & 62.87 & 45.60 & 64.50 & 62.64 & 1.00 & 89.50 \\
& CodeGPT & 62.75 & 63.64 & 47.28 & 65.32 & 63.08 & 1.50 & 89.50 \\
& UniXcoder & 61.27 & 62.25 & 46.35 & 64.45 & 62.94 & 1.50 & 94.50 \\
& CodeT5 & 63.69 & 64.36 & 46.02 & 65.96 & 63.27 & 1.50 & 97.00 \\
& {\tool}  & \textbf{66.53} & \textbf{67.18} & \textbf{52.03} & \textbf{68.52} & \textbf{65.22} & \textbf{2.50} & \textbf{100.00} \\
  \bottomrule
\end{tabular}
}
 \label{tab:RQ1}
\end{table*}

\begin{table}[h]
 \caption{The p-value between our proposed {\tool} and CodeT5 by using the Wilcoxon signed-rank test}
 \begin{center}
\begin{tabular}{c|cccccc}
  \toprule
\textbf{Corpus} & \textbf{$\mathit{BLEU}$} & \textbf{$\mathit{Weight}$-$\mathit{BLEU}$} & \textbf{$\mathit{Crystal}$-$\mathit{BLEU}$} & \textbf{$\mathit{Code}$-$\mathit{BLEU}$} & \textbf{$\mathit{Syntax}$-$\mathit{Match}$} \\
  \midrule
Lyra & 2.6e-3 & 1.7e-4 & 4.2e-2 & 5.8e-4 & 1.4e-3 \\
Pisces & 3.0e-10 & 6.5e-11 & 3.1e-16 & 4.3e-11 & 7.2e-5 \\
  \bottomrule
\end{tabular}
 \end{center}
 \label{tab:RQ1-pvalue}
\end{table}

Finally, we show two examples with the ground-truth code and the code generated by {\tool} and all baselines in the Lyra dataset and the Pisces dataset respectively. 
As shown in Fig.~\ref{fig:Exam-Python}, the code generated by CodeBERT contains syntax errors and fails to compile successfully. Additionally, the code generated by GraphCodeBERT deviates from the user's specifications as indicated by error messages. An exception in the SQL statement is present in the code generated by GPT, specifically an extraneous \code{The}. Furthermore, the code generated by CodeGPT and CodeT5 may contain a bug, i.e., the generated code lacks the return branch of \code{else}, it will return \code{jsonify(None)} during code execution resulting in an exception. 
In contrast, the code generated by {\tool} can accurately reflects the functional requirements, though differing from the ground-truth code in terms of naming identifiers.
As shown in Fig.~\ref{fig:Exam-Java}, the code generated by CodeBERT, GraphCodeBERT, GPT, and CodeGPT all have problems in their SQL statements. While the code generated by CodeT5 is semantically consistent with the ground-truth code, it diverges in terms of code style, specifically in the utilization of native SQL statements as opposed to the ORM framework employed by the ground-truth code. Therefore, the code generated by {\tool} shows higher similarity to the ground-truth code, both in semantics and code style.
Thus, we can observe that {\tool} can generate higher-quality code compared to  baselines.

\begin{tcolorbox}[width=1.0\linewidth, title={Summary for RQ1}]
{\tool} can significantly outperform the baselines in terms of six performance metrics on both Lyra and Pisces.
\end{tcolorbox}

\subsection{RQ2: What is the contribution of the multi-task learning of {\tool}?}
\label{sec:resultsRQ2}

\yg{To demonstrate the significance of each component in our proposed multi-task learning framework, we conduct a comparison of our approach with three of its variants:
}

\begin{itemize}
\item \textit{Variant\_1} does not introduce task-specific layers. This variant uses the same model and learns different tasks according to the prompt.

\item \textit{Variant\_2} does not introduce additional GLU for knowledge interaction. In this variant, the task-specific layers of the two tasks are separated and do not affect each other.

\item \yg{\textit{Variant\_3} does not introduce the auxiliary task and only consider the impact of the primary task. This variant model only contains CodeT5 with prompt and SF-beam search.}
\end{itemize}
\begin{figure*}[htbp]
\centering
\includegraphics[width=1.0\textwidth]{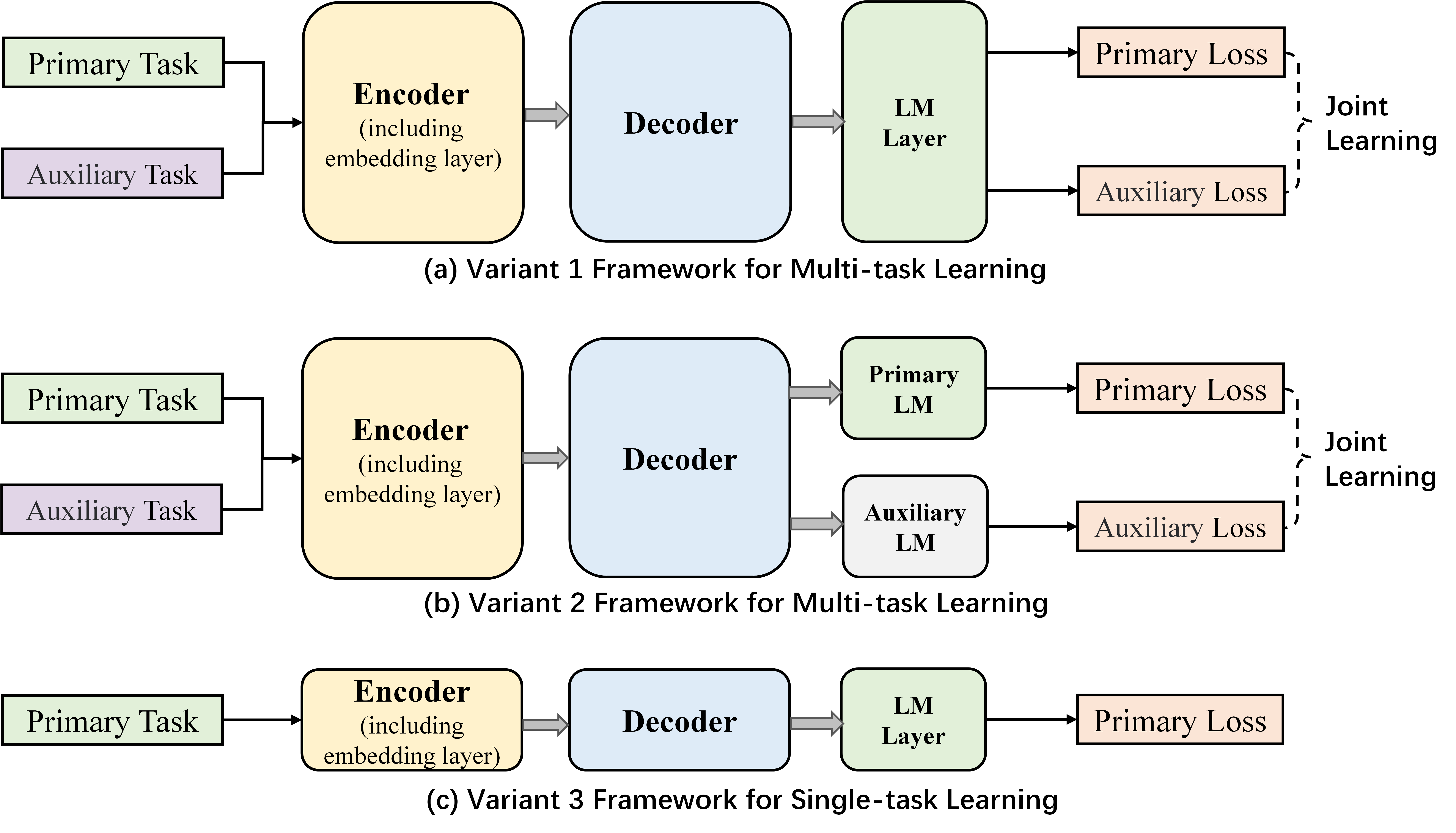}
\caption{The framework structure of the different variants}
\label{fig:Variants}
\end{figure*}

\begin{table*}[htbp]
\centering
 \caption{The comparison results between our proposed {\tool} and two variants}
\begin{tabular}{c|c|ccccccc}
  \toprule
\textbf{Corpus} & 
\textbf{Approach} & \textbf{$\mathit{BLEU}$} & \textbf{$\mathit{Weight}$-$\mathit{B}$} & \textbf{$\mathit{Crystal}$-$\mathit{B}$} & \textbf{$\mathit{Code}$-$\mathit{B}$} & \textbf{$\mathit{S}$-$\mathit{M}$} & \textbf{$\mathit{SE}$-$\mathit{M}$} & \textbf{$\mathit{C}$-$\mathit{E}$}\\
  \midrule
\multirow{3}{*}{Lyra}
& Variant\_1 & 77.67 & 78.86 & 69.11 & 80.96 & 85.45 & 29.00 & 99.00  \\
& Variant\_2 & 77.93 & 78.70 & 68.37 & 81.21 & 84.95 & 30.50 & 100.00  \\
& Variant\_3 & 76.97 & 77.32 & 68.10 & 81.03 & 84.25 & 30.50 & 97.50  \\
& {\tool}  & \textbf{78.36} & \textbf{79.23} & \textbf{69.77} & \textbf{82.72} & \textbf{86.92} & \textbf{34.00} & \textbf{100.00} \\
  \midrule
  \midrule
\multirow{3}{*}{Pisces} 
& Variant\_1 & 64.84 & 66.51 & 50.74 & 66.48 & 64.52 & 2.00 & 99.50  \\
& Variant\_2 & 65.85 & 65.95 & 51.59 & 67.24 & 64.48 & 2.00 & 100.00  \\
& Variant\_3 & 64.04 & 65.29 & 49.75 & 66.30 & 63.86 & 2.00 & 98.00  \\
& {\tool}  & \textbf{66.53} & \textbf{67.18} & \textbf{52.03} & \textbf{68.52} & \textbf{65.22} & \textbf{2.50} & \textbf{100.00} \\
  \bottomrule
\end{tabular}
 \label{tab:RQ2}
\end{table*}

\yg{In Table~\ref{tab:RQ2}, We can verify the effectiveness of task-specific layers and GLU by comparing {\tool} and variants. 
Similar to RQ1, we use $\mathit{BLEU}$, $\mathit{Weight}$-$\mathit{BLEU}$, $\mathit{Crystal}$-$\mathit{BLEU}$, $\mathit{Code}$-$\mathit{BLEU}$, ${Syntax}$-$\mathit{Match}$, ${SyntaxExact}$-$\mathit{Match}$, and $\mathit{Code}$-$\mathit{Executable}$ to evaluate the effectiveness of our proposed approach and these three variants.}
As shown in Table~\ref{tab:RQ2}, our proposed approach {\tool} shows superior performance compared to the other three variants on all metrics. Specifically, we can notice that when {\tool} removes only the GLU, there is a decrease in both lexical and syntactic similarity when compared to Variant 2 in Lyra and Pisces. On the contrary, Variant 1 shows comparable performance to that of Variant 2. 
\yg{When {\tool} removes the multitask learning framework and considers only the primary task, it can achieve lower performance than both variant 1 and variant 2 in terms of most metrics, especially on ${Syntax}$-$\mathit{Match}$ and $\mathit{Code}$-$\mathit{Executable}$.}
These findings suggest that, within a multi-task learning framework, the inclusion or exclusion of task-specific layers does not have a significant impact on performance if there is no additional semantic interaction. 
\yg{Furthermore, we show the importance of multitask learning by comparing it with variant 3, which retains only prompt and SF-beam search compared to {\tool}, but is also can outperform CodeT5 with $\mathit{Code}$-$\mathit{Executable}$, which shows that the construction of prompt and compiler-based SF-beam search are also beneficial for CodeT5.}
Moreover, these results indicate that the incorporation of the GLU in {\tool} effectively embeds the grammatical knowledge acquired through the auxiliary task into the primary task, thereby improving the model's performance.

\begin{tcolorbox}[width=1.0\linewidth, title={Summary for RQ2}]
Different from previous multi-task learning frameworks, we incorporate GLU after task-specific layers, which enables the incorporation of knowledge acquired through the auxiliary task into the primary task. Our study shows this setting can result in a positive impact on the performance of {\tool}.
\end{tcolorbox}

\subsection{RQ3: What is the benefit of using the SF-Beam Search method of {\tool}?}
\label{sec:resultsRQ3}

\begin{table*}[h]
 \caption{The comparison results between SF-Beam Search and other decoding algorithms}
     \centering
\begin{tabular}{c|c|ccccccc}
  \toprule
\textbf{Corpus} & 
\textbf{Approach} & \textbf{$\mathit{BLEU}$} & \textbf{$\mathit{Weight}$-$\mathit{B}$} & \textbf{$\mathit{Crystal}$-$\mathit{B}$} & \textbf{$\mathit{Code}$-$\mathit{B}$} & \textbf{$\mathit{S}$-$\mathit{M}$} & \textbf{$\mathit{SE}$-$\mathit{M}$} & \textbf{$\mathit{C}$-$\mathit{E}$}\\
  \midrule
\multirow{4}{*}{Lyra}
& Sampling Search & 76.26 & 76.91 & 65.38 & 79.55 & 83.23 & 29.00 & 94.50  \\
& Greedy Search & 77.89 & 78.75 & 69.12 & 81.07 & 84.86 & 31.50 & 96.50  \\
& Beam Search & 78.03 & 78.94 & 69.17 & 81.44 & 86.08 & 32.50 & 97.00  \\
& SF-Beam Search  & \textbf{78.36} & \textbf{79.23} & \textbf{69.77} & \textbf{82.72} & \textbf{86.92} & \textbf{34.00} & \textbf{100.00} \\
  \midrule
  \midrule
\multirow{4}{*}{Pisces} 
& Sampling Search & 64.52 & 64.48 & 49.48 & 66.23 & 63.98 & 1.50 & 96.50  \\
& Greedy Search & 65.12 & 66.04 & 50.98 & 66.94 & 64.50 & 2.00 & 97.00  \\
& Beam Search & 66.05 & 66.15 & 51.86 & 67.36 & 64.98 & 2.00 & 98.00  \\
& SF-Beam Search & \textbf{66.53} & \textbf{67.18} & \textbf{52.03} & \textbf{68.52} & \textbf{65.22} & \textbf{2.50} & \textbf{100.00} \\
  \bottomrule
\end{tabular}
 \label{tab:RQ3}
\end{table*}

To investigate how the SF-Beam Search method affects the performance of {\tool}, we mainly consider three different decoding algorithms:

\begin{itemize}
\item \textit{Sampling Search.} It is a sampling algorithm with randomness. Compared to the algorithm by probability, this sampling algorithm can introduce more randomness and is often present in dialogue generation task~\citep{liu2020you}.

\item \textit{Greedy Search.} It is a simple yet effective sampling algorithm that directly selects the token with the maximum probability for each output until a terminator appears or the maximum sentence length is reached. 

\item \textit{Beam Search.} It is a heuristic graph search algorithm that keeps top-$k$ token nodes with the highest probability at each step, which can reduce the space and time cost occupied by the search.

\end{itemize}

We show the comparison results in Table~\ref{tab:RQ3} and we find that the SF-Beam Search method works best both on Lyra and Pisces. 
Compared with Sampling Search, Greedy Search, and Beam Search, our SF-Beam Search method outperforms them in all metrics, especially in code execution rate.
Therefore, the results show the effectiveness of our proposed GE-BS method.

Furthermore, as our SF-Beam Search method employs the beam search technique, the size of the beam size can have a direct impact on the performance of code generation. Thus, we conduct a comparison of the BLEU metric trends for code generated by the SF-Beam Search method and the beam search method for different beam sizes.

\yg{Later, we conduct a time cost analysis when considering program compiler cost during beam search in our SF-Beam Search method. 
Specifically, we find that without optimization, the average time required for the program compiler is approximately 1$\sim$2 seconds. This additional time cost resulted in a total time cost of approximately 15 seconds for each test sample when the beam search is set to 10. However, by using the multi-threading technology, the time cost for each test sample can be reduced to around 2$\sim$3 seconds when the beam search is 10.}

\begin{figure}[htbp]
	\centering
	\subfigure[The hyper-parameter beam size in Lyra]{%
		\includegraphics[width=0.45\textwidth]{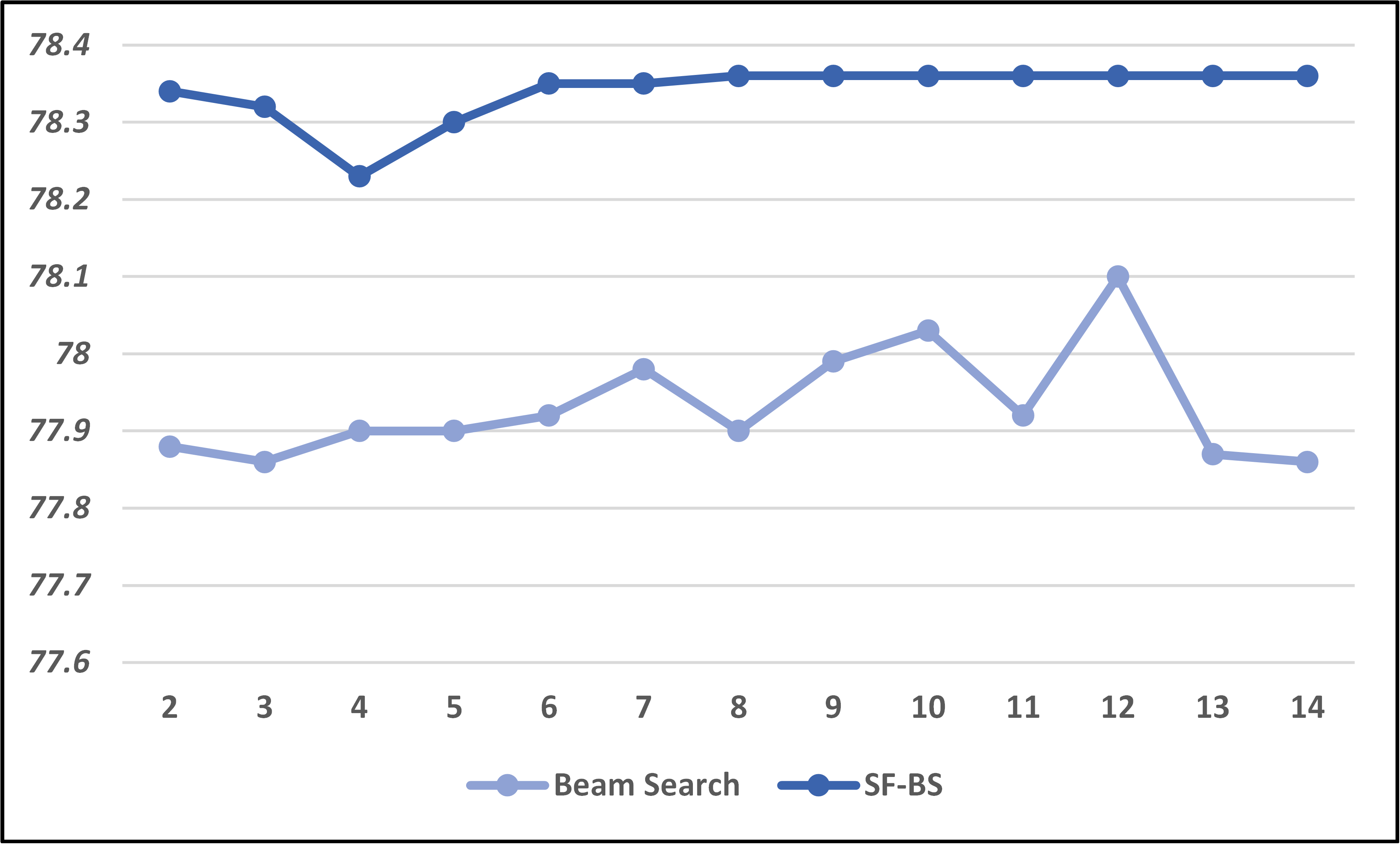}%
	}%
        \hspace{2mm}
	\subfigure[The hyper-parameter beam size in Pisces]{%
		\includegraphics[width=0.45\textwidth]{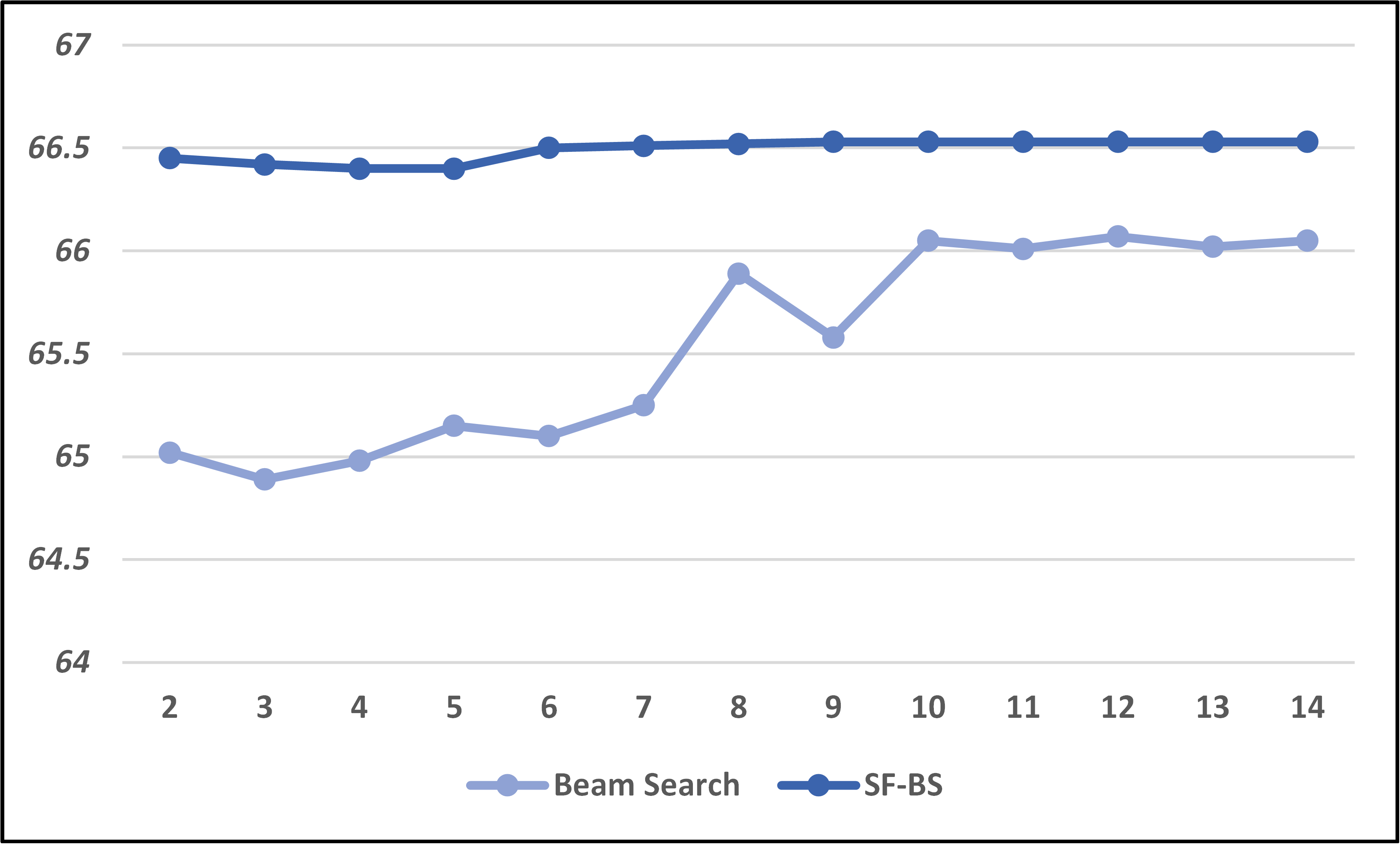}%
	}%
	\caption{The influence of the hyper-parameter beam size in Lyra and Pisces, where the horizontal axis denotes beam size and the vertical axis denotes the value of BLEU}
	\label{fig:beam_size}
	\vspace{-1mm}
\end{figure}

Finally, we conduct a series of experiments for the influence of the hyper-parameter beam size. As shown in Fig.~\ref{fig:beam_size}, we can observe that the performance of code generated by the beam search method exhibits fluctuations as the beam size increases, while the code generated by the SF-Beam Search method demonstrates higher stability. Furthermore, the quality of code generated by the SF-Beam Search method is higher than that of the beam search method when analyzing all the results.

\begin{tcolorbox}[width=1.0\linewidth, title={Summary for RQ3}]
In comparison to other decoding algorithms, the utilization of SF-Beam Search can result in a positive impact on the performance of Turducken-style code generation and demonstrates higher stability.
\end{tcolorbox}

\subsection{RQ4: What is the impact of different prompt methods of {\tool}?}
\label{sec:resultsRQ4}

\yg{In our study, we manually define the natural language tokens in hard prompt templates. To explore the impact of hard prompts and other prompt methods on {\tool}, we analyzed the performance of different hard prompts, soft prompts, and mixed prompts in terms of seven automatic metrics in this RQ:}

\begin{itemize}
\item \textit{[X] [Y]} means that the hard prompt is not used and relies only on the final LM layer for judgment.

\item \textit{[TASK] : [X] [Y]} means to use the name of the task as the hard prompt, without adding any other prompt token.

\item \textit{Generate Turducken-Style code under [TASK] : [X] [Y]} has the same meaning as the prompt used in {\tool}, but we add some tokens to make the prompt longer.

\item \yg{\textit{[SOFT] $\ast$ n : [X] [Y]} means to prepend several virtual tokens to the original input, which is referred as prefix-tuning~\citep{li2021prefix} and is a typical method of soft prompt~\citep{wang2022no}.} 

\item \yg{\textit{[SOFT] $\ast$ n + \textit{Generate [TASK] code : [X] [Y]}} means adding several virtual tokens before the hard prompt, which is referred as mixed prompt~\citep{longpre2023flan}.} 

\end{itemize}

\begin{table*}[h]
 \caption{The comparison results between different prompts}
     \centering
\scalebox{0.9}{
\begin{tabular}{c|c|ccccccc}
  \toprule
\textbf{Corpus} & 
\textbf{Approach} & \textbf{$\mathit{BLEU}$} & \textbf{$\mathit{Weight}$-$\mathit{B}$} & \textbf{$\mathit{Crystal}$-$\mathit{B}$} & \textbf{$\mathit{Code}$-$\mathit{B}$} & \textbf{$\mathit{S}$-$\mathit{M}$} & \textbf{$\mathit{SE}$-$\mathit{M}$} & \textbf{$\mathit{C}$-$\mathit{E}$}\\
  \midrule
\multirow{8}{*}{Lyra}
& \textit{[X] [Y]} & 78.16 & 78.89 & 69.54 & 82.45 & 84.97 & 29.00 & \textbf{100.00}  \\
& \textit{[TASK] : [X] [Y]} & 77.68 & 78.45 & 68.90 & 81.92 & 84.06 & 33.00 & 99.50  \\
& \textit{\tabincell{c}{Generate Turducken-Style code \\ under [TASK] : [X] [Y]}} & 78.28 & 79.06 & 69.16 & 82.57 & 85.42 & \textbf{34.00} & 99.00  \\
& \textit{[SOFT] $\ast$ n : [X] [Y]} & 78.19 & 79.08 & 69.49 & 82.47 & 85.04 & 32.00 & 99.50  \\
& \textit{\tabincell{c}{[SOFT] $\ast$ n + \\ Generate [TASK] code : [X] [Y]}} & 78.32 & 79.24 & 69.69 & \textbf{82.73} & 86.78 & \textbf{34.00} & \textbf{100.00}  \\
& \textit{Generate [TASK] code : [X] [Y]} & \textbf{78.36} & \textbf{79.23} & \textbf{69.77} & 82.72 & \textbf{86.92} & \textbf{34.00} & \textbf{100.00} \\
  \midrule
  \midrule
\multirow{8}{*}{Pisces} 
& \textit{[X] [Y]} & 65.84 & 66.23 & 51.48 & 67.93 & 64.66 & 2.00 & \textbf{100.00}  \\
& \textit{[TASK] : [X] [Y]} & 65.26 & 65.89 & 51.38 & 67.58 & 64.44 & 2.00 & \textbf{100.00}  \\
& \textit{\tabincell{c}{Generate Turducken-Style code \\ under [TASK] : [X] [Y]}} & 66.04 & 66.76 & 51.23 & 67.94 & 64.50 & \textbf{2.50} & 99.50  \\
& \textit{[SOFT] $\ast$ n : [X] [Y]} & 65.28 & 66.15 & 51.44 & 67.87 & 64.51 & 2.00 & 99.00  \\
& \textit{\tabincell{c}{[SOFT] $\ast$ n + \\ Generate [TASK] code : [X] [Y]}} & \textbf{66.74} & 67.12 & \textbf{52.09} & 68.12 & 65.12 & \textbf{2.50} & 99.00  \\
& \textit{Generate [TASK] code : [X] [Y]} & 66.53 & \textbf{67.18} & 52.03 & \textbf{68.52} & \textbf{65.22} & \textbf{2.50} & \textbf{100.00} \\
  \bottomrule
\end{tabular}
}
 \label{tab:RQ4}
\end{table*}

\yg{We follow the setting of Wang et al.~\citep{wang2022no} and set the value of the parameter $n$ in the soft prompts and mixed prompts to 4.
We show the comparison results in Table~\ref{tab:RQ4}. 
In this table, we can find that the different prompts have a slight impact on the performance, and using hard prompt templates for {\tool} can achieve the best performance on both Lyra and Pisces in terms of most metrics.
More importantly, a well-designed hard prompt template is particularly important. A good prompt can stimulate the potential of the model, but this may require the experts to rely on their own experience for tuning.
Moreover, our results indicate a similarity in the model performance between the incorporation of solely the soft prompt and the absence of any prompt. However, the introduction of a mixed prompt can lead to optimal model performance in several metrics. This result shows the significance of the hard prompt and the possibility of improving model performance through the implementation of the soft prompt as a precursor to a well-designed hard prompt.
}

\begin{tcolorbox}[width=1.0\linewidth, title={Summary for RQ4}]
We demonstrate the importance of different prompt methods and find only well-designed prompts can have a positive impact on the performance of {\tool}.
\end{tcolorbox}
\section{Human Evaluation}
\label{sec:discussion}

In RQ1, we conducted performance comparisons automatically in terms of six performance metrics. 
However, in the absence of test cases, these automatic performance metrics may not truly reflect the semantic similarity between different code snippets.
To alleviate this issue, we further conducted a human evaluation to verify the effectiveness of our proposed approach {\tool}. 
In our human study, we only compare {\tool} with CodeT5, which can achieve the best performance in all baselines.

We refer to the methodology used by \citep{DBLP:conf/icse/HuX0WCZ22} and \citep{yang2022exploitgen} to conduct the human evaluation in the  code generation task. 
In the human study, we evaluate the quality of the generated code from three aspects:

\begin{itemize}
\item \texttt{Code Readability.} It evaluates the code readability of the generated code. For example, Java code should follow DRAFT\footnote{\url{http://cr.openjdk.java.net/~alundblad/styleguide/index-v6.html}} and Python code should follow PEP 8\footnote{\url{https://peps.python.org/pep-0008}}.

\item \texttt{Semantic Similarity.} It evaluates the semantic similarity between the generated code and the reference code since the code snippets with the same semantics may differ at the lexical level.

\item \yg{\texttt{User Preference.} It evaluates the users' preference score between the reference code and code generated by the two methods, which can represent the real preferences of the users and avoid the bias caused by the provided ground truth.}
\end{itemize}

The scores of \texttt{Code Readability} and \texttt{Semantic Similarity} range from 0 to 4 (the higher the better) and all these scores are integers. 
\yg{In consideration of \texttt{User Preference}, we designed it as a choice question, where volunteers were asked to select the one or more codes they think best from three candidate codes (i.e., the reference code, the code generated by CodeT5, and the code generated by TurduckenGen).}
We invite five volunteers, who have 3$\sim$5 years of Java and Python experience and have good English reading ability. 
Then we paid a certain fee for each participant in their code evaluation work.
Due to the high cost of manually analyzing all these samples in the testing set, we randomly select 100 Turducken-style code snippets and functional descriptions by following \citep{9678552}.
Specifically, we randomly selected 50 samples from the Lyra and 50 samples from the Pisces.
\yg{To avoid any potential bias (i.e., answer leakage) caused by previously sampled individuals in the Code Readability and Semantic Similarity stages, we randomly selected new 50 samples for the User Preference questionnaire for the same group of volunteers.}


\begin{figure}[]
\centering
\includegraphics[width=0.85\textwidth]{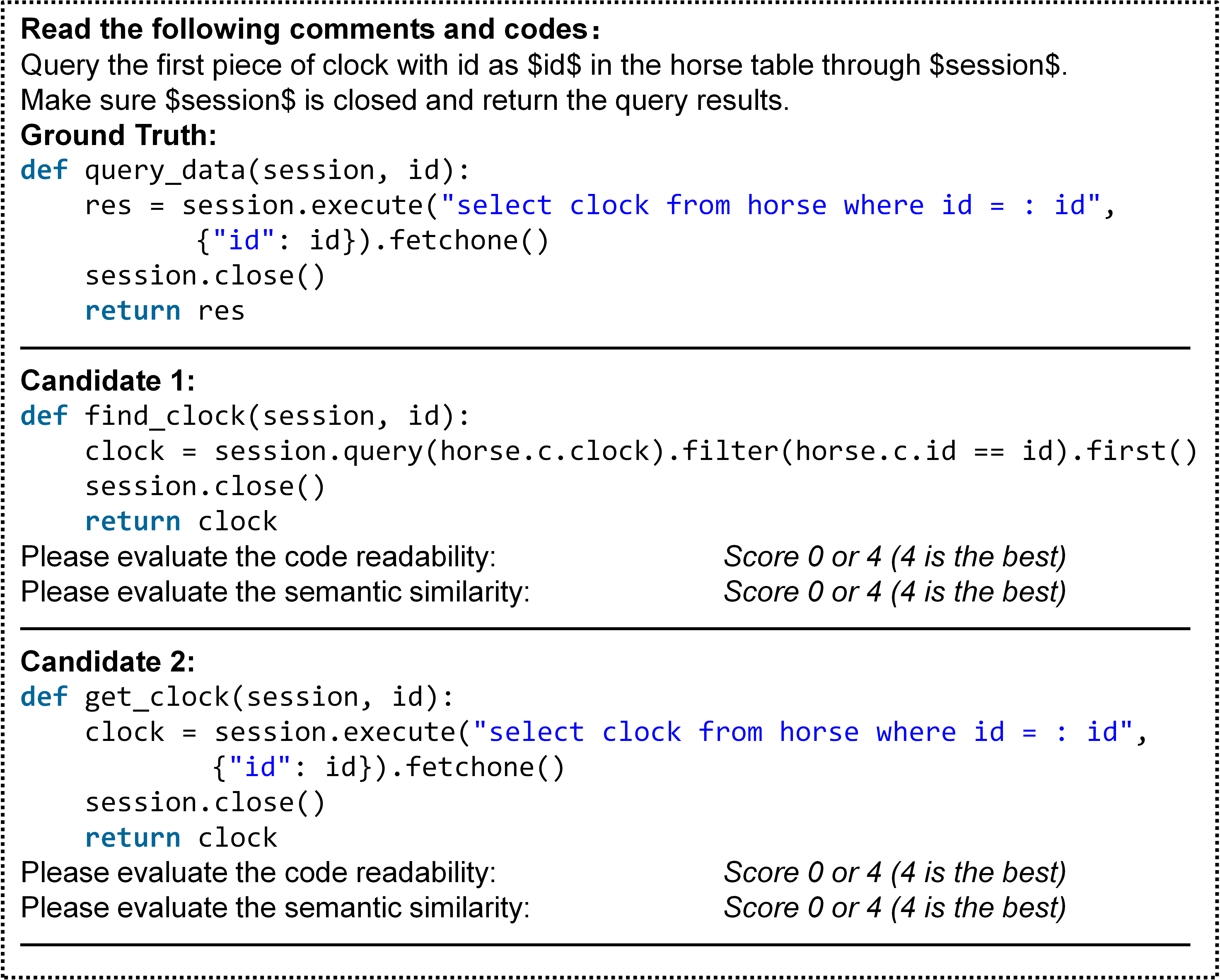}
\caption{A questionnaire used to evaluate Code Readability and Semantic Similarity in our human evaluation}
\label{fig:structure}
\end{figure}

\begin{figure}[]
\centering
\includegraphics[width=0.9\textwidth]{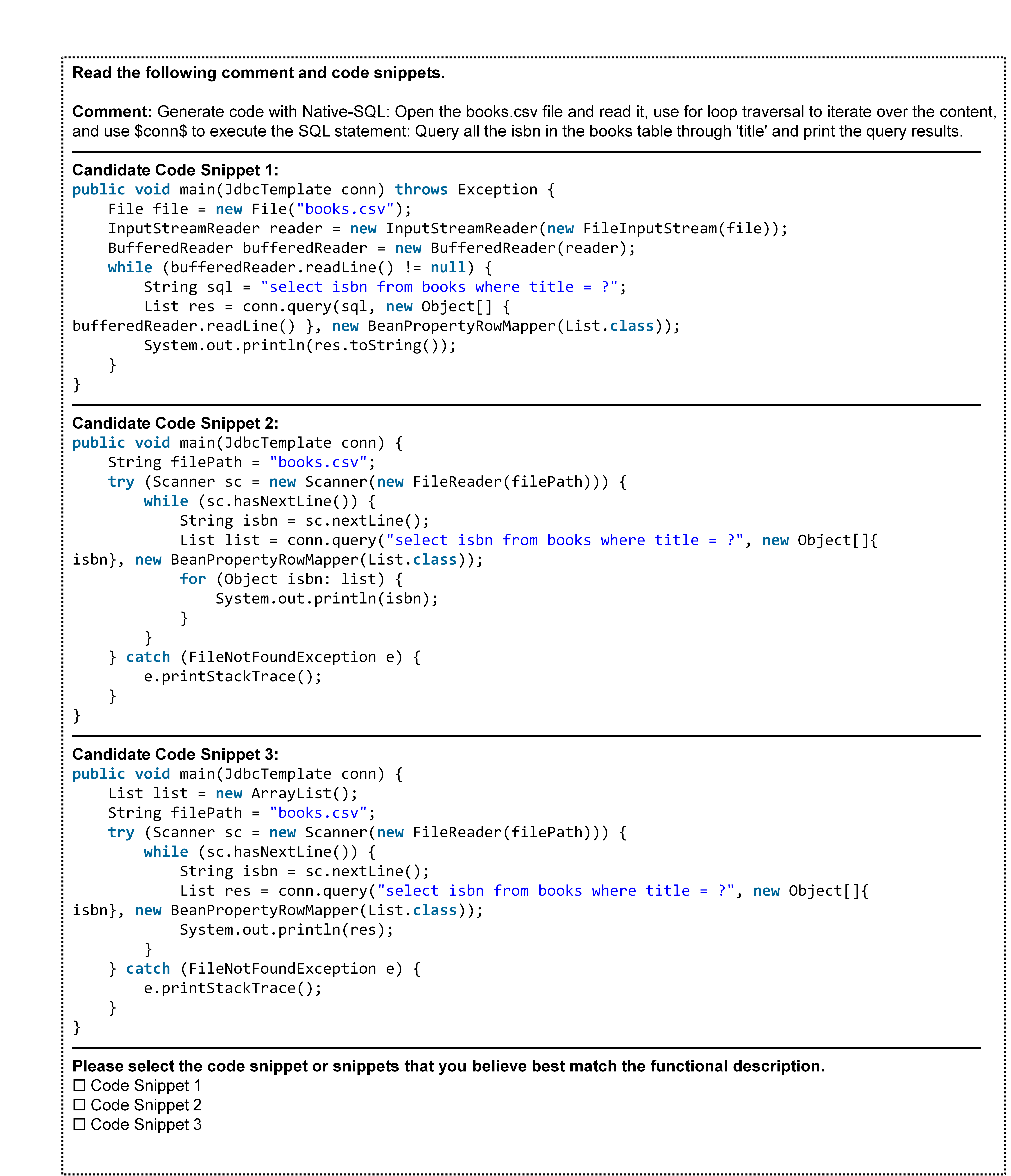}
\caption{A questionnaire used to evaluate User Preference in our human evaluation}
\label{fig:structure-2}
\end{figure}

For each code snippet, we generate a questionnaire for each participant, which is shown in Fig.~\ref{fig:structure} and Fig.~\ref{fig:structure-2}. 

In the questionnaire used to evaluate Code Readability and Semantic Similarity, there are two code snippets generated by {\tool} and the baseline CodeT5 respectively. 
Each participant is asked to score each code in terms of code readability and semantic similarity for two code snippets generated by {\tool} and the baseline CodeT5 respectively. 
\yg{In the questionnaire used to evaluate user preference, there are three code snippets, which contain the reference code, the code generated by {\tool}, and the code generated by CodeT5 respectively. 
Since there is the possibility of the exact match between these three code snippets, each participant is asked to choose one or more codes that best match the functional description.}

During the code quality evaluation process, the participants can discuss and resort to external resources (e.g., Wikipedia and Q\&A websites). 
To ensure the fairness of the comparison, the participants do not know which code is generated by which approach, and the order of questionnaires is different for different participants. 
To guarantee the code evaluation quality, we need each participant to review only 25 code snippets in half a day to avoid fatigue. 

\begin{table}[]
  \begin{center}
 \caption{Results of our human study in terms of code readability, semantic similarity (values in parentheses indicate Kendall's coefficient of concordance in our human evaluation), and user preference}
 \label{tab:human}
\begin{tabular}{c|ccccc}
\toprule
\textbf{Dataset} & \textbf{Approach} & \textbf{Code Readability} & \textbf{Semantic Similarity} & \textbf{User Preference}\\
\midrule
\multirow{3}{*}{Lyra} & Reference   & -  & -  &  86.6\% \\
& CodeT5   & 3.152 (0.533)    & 3.136 (0.788) & 52.4\%  \\
& {\tool} & \textbf{3.492} (0.553) &  \textbf{3.652} (0.843) & 68.5\%  \\
\midrule
\multirow{3}{*}{Pisces} & Reference   & -  & -  & 90.4\%  \\
& CodeT5   & 2.464 (0.771)  & 2.472 (0.884)  & 56.2\%  \\
& {\tool} & \textbf{2.888} (0.666)  &  \textbf{3.192} (0.756)  & 64.7\% \\
\bottomrule
\end{tabular}
 \end{center}
\end{table}

\begin{figure}[]
\centering
\includegraphics[width=0.75\textwidth]{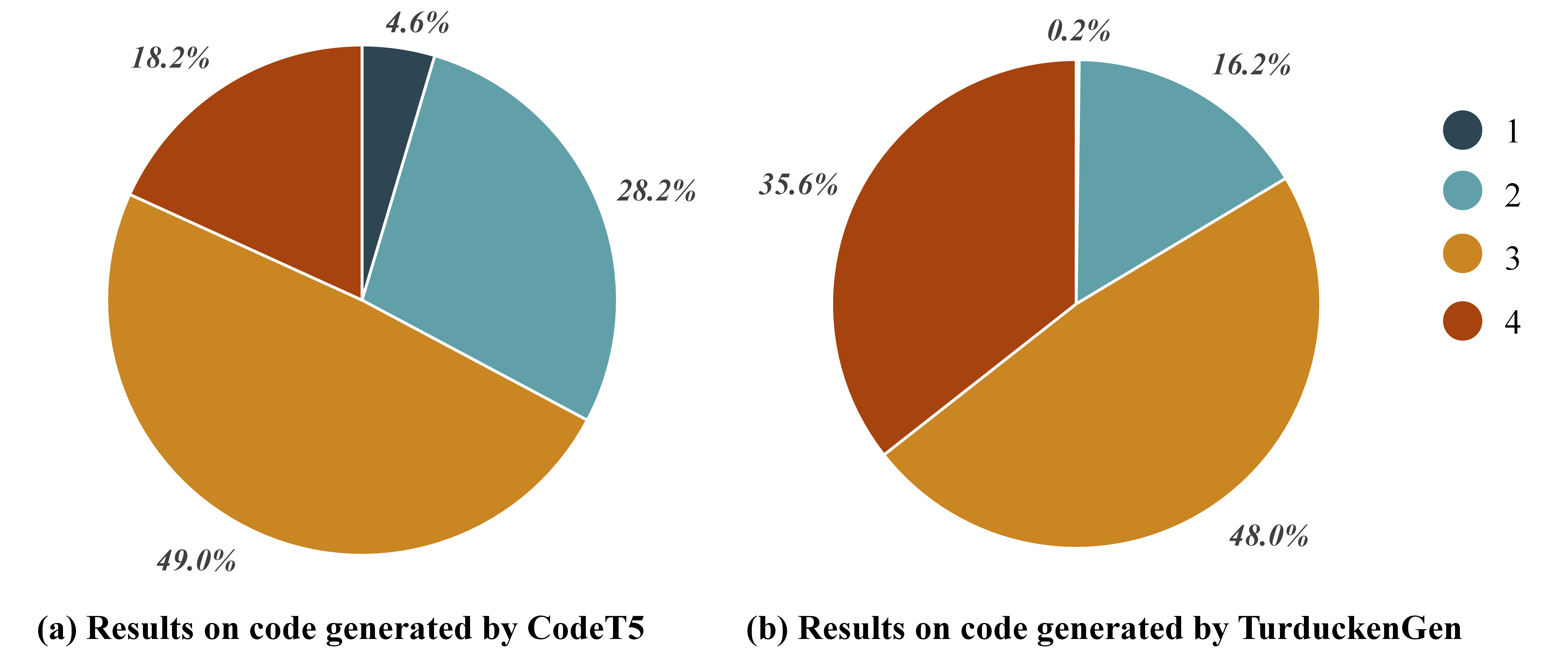}
\caption{Rating distribution of human evaluation in terms of code readability}
\label{fig:read}
\end{figure}

\begin{figure}[]
\centering
\includegraphics[width=0.75\textwidth]{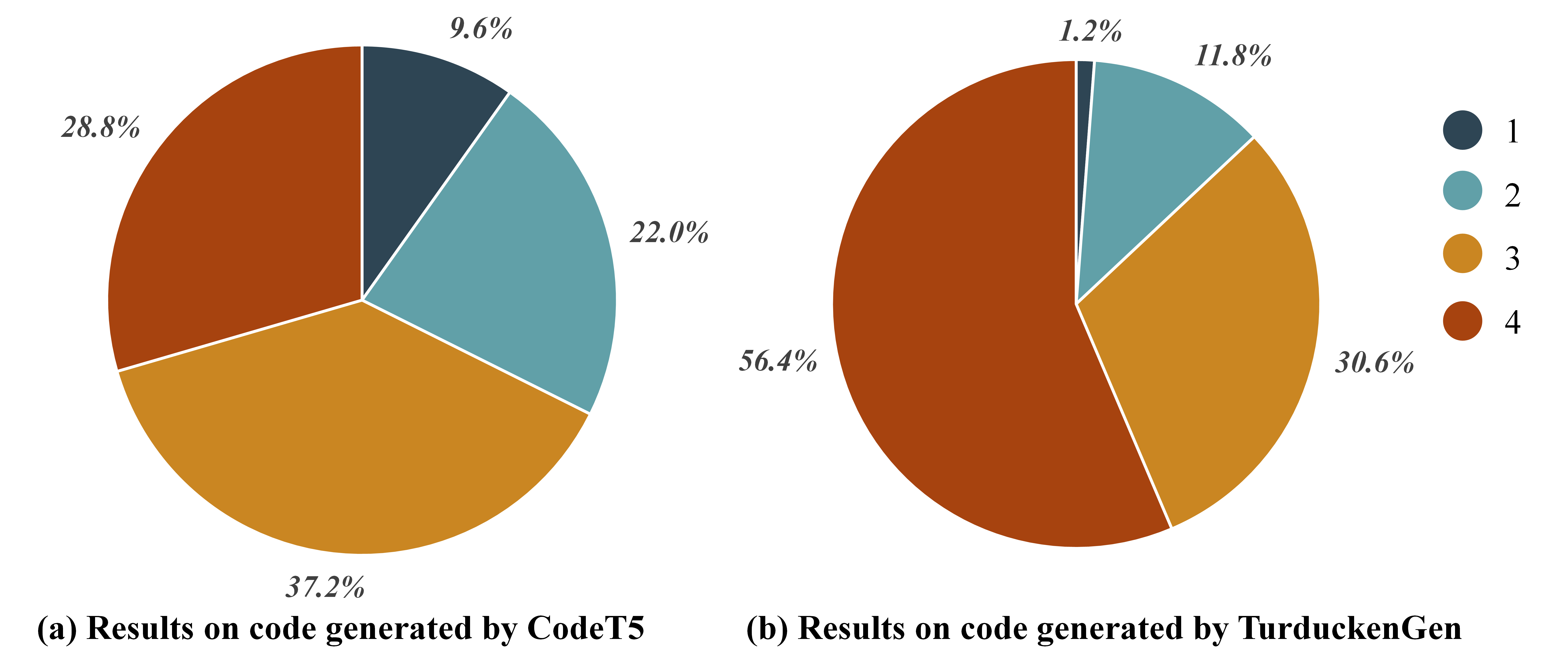}
\caption{Rating distribution of human evaluation in terms of semantic similarity}
\label{fig:Semantic}
\end{figure}

Fig.~\ref{fig:read} and Fig.~\ref{fig:Semantic} show the statistical results of this feedback in terms of code readability and semantic similarity. 
In these two figures, the left and right sub-figures show the votes for the Turducken-style code generated by CodeT5 and {\tool}, respectively. 
In terms of code readability, we find that 83\% of the human ratings for the code generated by {\tool} are not less than 3, which means that they are considered to have good programming specifications and can be easily comprehended by the user.
In terms of semantic similarity, we find that 87\% of the human ratings for the code generated by {\tool} are not less than 3, which means they are considered to have good quality and can be used as code without major modifications.

We also compute the average score of the participants' feedback and the results are shown in Table~\ref{tab:human}. In this table, values in parentheses indicate the degree of consistency of the scores marked by the participants using Kendall’s coefficient of concordance~\citep{legendre2005species}. 
Notice the value of Kendall's coefficient of concordance ranges from 0 to 1, with larger values indicating higher concordance.
Based on the results, we can find that for the Lyra, {\tool} can outperform the approach CodeT5 by 0.340 and 0.516 respectively in terms of code readability and semantic similarity. 
For example, in the corpus Pisces, {\tool} can outperform the approach CodeT5 by 0.424 and 0.720 respectively in terms of code readability and semantic similarity.
In addition, all Kendall's coefficient of concordance in our human evaluation is larger than 0.5, which indicates their scores are about the same degree of concentration. 

\yg{As for user preference, we first calculate the score for each volunteer, which is the proportion of times they chose the reference code, the code generated by CodeT5, and the code generated by {\tool} over all questionnaires. Then, we average the scores of all volunteers to obtain the final score. Based on the results, we can find that regardless of Lyra or Pisces, volunteers tend to prefer the reference code. Secondly, volunteers generally believe that the code generated by {\tool} is better than the code generated by CodeT5, indicating that volunteers are more inclined towards the code generated by {\tool}.}
Therefore, our human study can further verify the competitiveness of {\tool}.

\section{Threats to Validity}
\label{sec:threats}

\subsection{Construct Validity}

The first construct validity concerns the appropriateness of our evaluation metrics. To mitigate this construct threat, we consider six evaluation metrics. Additionally, we also compute the $p$-value by utilizing the Wilcoxon signed-rank test to further verify the statistical significance of our proposed approach. 
The second construct is the semantic correctness of  the code generated by {\tool}.
Due to the lack of high-quality test cases, 
we mainly conduct a human evaluation and assess the effectiveness of {\tool} by taking into account the code readability and semantic similarity of the generated Turducken-style code.

\subsection{Internal Validity}

The first internal threat is the potential defects in the implementation of our proposed method. 
To alleviate this threat, we first check the code carefully and use mature libraries, such as PyTorch and Transformers.
The second internal threat is the implementation of the baseline methods. To alleviate this threat, we try our best to fine-tune the pre-trained models (i.e., CodeBERT\footnote{\url{https://huggingface.co/microsoft/codebert-base}}, GraphCodeBERT\footnote{\url{https://huggingface.co/microsoft/graphcodebert-base}}, GPT\footnote{\url{https://huggingface.co/gpt2}}, CodeGPT\footnote{\url{https://huggingface.co/microsoft/CodeGPT-small-java}}\footnote{\url{https://huggingface.co/microsoft/CodeGPT-small-py}}, and CodeT5\footnote{\url{https://huggingface.co/Salesforce/codet5-base}}).
\yg{Moreover, the parameters for both CodeT5 and {\tool} are the same, as shown in Table~\ref{Hyper-parameters}. For the other baseline models, we followed the parameter setting of the previous study~\citep{liang2021lyra}.}
\yg{The third internal threat is that our proposed architecture is specifically designed for pre-trained encoder-decoder models, such as CodeT5. Our approach is not applicable to other pre-trained models, such as CodeBERT, GPT, GraphCodeBERT, etc. This is because CodeBERT-like models are only pre-trained on the encoder side, while our multi-task learning approach operates on the decoder side. On the other hand, GPT-like models are only pre-trained on the decoder side, while the prompt method proposed by our approach relies more heavily on the encoder side. Therefore, we believe that our method is only suitable for pre-trained encoder-decoder models, such as CodeT5. }

\subsection{External Validity}

\yg{External validity refers to the extent to which the findings of a study can be generalized to other datasets. In our study, we mainly focused on Turducken-style code generation. 
The main difference between generating Turducken-style code and general code is the nested nature of Turducken-style code (i.e. declarative programs are embedded in imperative programs). In addition, it is more common for programmers to write Turducken-style code in real business development scenarios. 
However, our approach is essentially independent of specific programming languages, meaning that it has a certain degree of generality. By adjusting the training data and fine-tuning the model, our proposed approach can be applied to general code generation tasks. Meanwhile, the corresponding compiler needs to be replaced for a specific programming language in order to parse the AST and perform the SF-Beam Search decoding.}

\section{Related Work}
\label{sec:related}

In this section, we mainly summarize related studies for automatic code generation, multi-task learning for code intelligence, and abstract syntax tree traversal methods. After related work analysis, we emphasize the novelty of our study.

\subsection{Automatic Code Generation}

In the field of automatic code generation for declarative programming, most researchers focused on the automatic generation of SQL statements. 
\citep{dahl1994expanding} were the first to investigate this area and shared the ATIS dataset for the Airline Travel Query domain. Then, a growing number of researchers explored and constructed various datasets for automatic SQL generation. For example, some datasets~\citep{zelle1996learning, iyer2017learning} focus on a single domain and other datasets~\citep{zhong2017seq2sql, yu2018spider, yu2019sparc, yu2019cosql} encompass multiple domains and cover diverse SQL types. 
Early studies~\citep{popescu2003towards, mahmud2015rule} primarily employed rule-based approaches. However, these methods required pre-defined templates for SQL statements and had poor scalability. With the advent of neural network models, the majority of work in this field has shifted towards generating more flexible SQL statements with a wider range of styles. \citep{yu2018spider} modeled SQL generation as a neural translation task and employed an RNN-based Seq2Seq model for this purpose. \citep{yu2018typesql} additionally incorporated Schema Linking in the encoder and added Sketch information in the decoder. \citep{bogin2019representing} then used GNN as the encoder and LSTM as the decoder and incorporated grammar information. 
With the emergence of Transformer~\citep{vaswani2017attention}, more researchers~\citep{wang2020rat, lin2020bridging, huang2021relation, rubin2021smbop} applied this technique to SQL generation tasks and considered domain-specific language information to improve performance. 
Recently, inspired by pre-trained language models, \citep{scholak2021picard} applied the T5 model and achieved state-of-the-art performance.

For the automatic code generation of imperative programming, \citep{mou2015end} were the first to propose the use of a standard encoder-decoder architecture based on RNN to generate corresponding C++ code snippets from user functional descriptions. \citep{ling2016latent} proposed two new corpora for card games code and employed the LSTM to construct the encoder-decoder architecture, which incorporates character-level softmax and pointer networks. \citep{yin2018tranx} proposed TRANX, a model based on ASDL, LSTM, the attention mechanism, and the copy mechanism for Python code generation. \citep{hayati2018retrieval} proposed RECODE, which is based on TRANX and integrates information retrieval methods into the Seq2Seq model. \citep{wei2019code} and \citep{DBLP:conf/wcre/YangCZY22} formalized the code generation task and code summarization task as dual tasks and improved model performance through dual learning. \citep{sun2019grammar} designed a grammar-based structural convolutional neural network for code generation that generated a program by predicting the programming language's grammar rules and subsequently proposed a novel tree-based neural architecture TreeGen~\citep{sun2020treegen} based on Transformer. 
With the development of pre-trained language models for code-related tasks, \citep{liguori2021evil} and \citep{yang2022exploitgen} applied CodeBERT to the exploit code generation task. \citep{DBLP:conf/nips/LuGRHSBCDJTLZSZ21} proposed CodeGPT, \citep{ahmad2021unified} proposed PLBART, and \citep{wang2021codet5} proposed CodeT5 for improved Java code generation task. 
Recently, researchers~\citep{wang2022compilable, le2022coderl} combined pre-trained language models and reinforcement learning to optimize pre-trained language models from the perspective of code compilability.

For the automatic Turducken-style code generation, to our best knowledge, there was only one study~\citep{liang2021lyra}. They used Encoder-Only pre-trained models and Decoder-Only pre-trained models to conduct experiments on the Lyra dataset to demonstrate the potential possibility of Turducken-style code generation.

In contrast to previous studies, our goal is to generate syntactically correct Turducken-style code, and we consider the learning paradigm of multi-task learning to construct ingenious auxiliary tasks. Moreover, we propose a syntax-first decoding algorithm to satisfy the syntax of the generated code.

\subsection{Multi-task Learning for Code Intelligence}

Multi-task learning (MTL) is a machine learning technique that allows a model to learn multiple tasks simultaneously by sharing common features or representations. In the field of code intelligence, multi-task learning has been applied to various tasks, such as code understanding, code summarization, and code completion. 
\citep{wang2021mulcode} proposed a multi-task learning approach MulCode for source code understanding. This approach learns unified representation space for tasks,  with the pre-trained BERT model for the token sequence and the Tree-LSTM model for abstract syntax trees. MulCode achieves promising performance on three downstream tasks: comment classification, author attribution, and duplicate function detection.
For the code summarization task, \citep{xie2021exploiting} proposed a deliberation multi-Task learning approach DMACOS by exploiting method names to improve code summarization. They introduced the tasks of generation and informativeness prediction of method names as two auxiliary training objectives for code summarization. Then they incorporated a novel two-pass deliberation mechanism into their MTL architecture.
For the code completion task, \citep{liu2020multi} and \citep{liu2022unified} adopted multi-task learning to predict the token and its type jointly and utilize the predicted type to assist the token prediction.
In summary, multi-task learning can be used to improve the performance of code intelligence tasks by allowing models to learn multiple tasks and share common features or representations.

Compared with previous multi-task learning studies that were used for code intelligence, our approach designs a syntax-guided auxiliary task using for code-based generation and uses a GLU network to incorporate the syntax knowledge learned from the auxiliary task into the primary task. Empirical results show the promising performance of our proposed approach.

\subsection{Abstract Syntax Tree Traversal Methods}

An Abstract Syntax Tree (AST) is a tree representation of the abstract structure of source code written in a programming language.
Researchers usually use AST to extract syntactic knowledge of the code. 
In the field of code comment generation, \citep{yang2021comformer} proposed Sim\_SBT, which uses pre-order traversal to flatten the type nodes of the AST.
\citep{hu2020deep} and \citep{niu2022spt} proposed SBT and X-SBT, respectively. They also used pre-order traversal, along with brackets and xml-like tags to form structured data.
In the field of code translation, \citep{liu2023syntax} captured the structural information of the AST by modeling the node distances and the node paths.

Different from previous AST traversal methods, our proposed SAT traversal method takes into account both the type and value of nodes in the AST. By incorporating the type information of the parent node into the code tokens explicitly, using our proposed SAT traversal method can help to generate code, which satisfy syntactic constraints.

\section{Conclusion and Future Work}
\label{sec:conclusion}

In this study, we propose a novel approach {\tool} for Turducken-style code generation, where {\tool} addresses the challenge of syntactic constraints on code generation from three perspectives. 
The results of the automated evaluation show that our proposed {\tool} outperforms the state-of-the-art baselines  on Lyra and Pisces. 
Moreover, we also verify our SF-Beam Search decoding method and model components rationality of {\tool} by designing a set of ablation studies. 
Finally, We conduct a human study to evaluate the quality of the generated code in terms of code readability and semantic similarity from the practitioner's perspective.

In the future, we first want to further improve the performance of {\tool} by considering advanced code representation methods. We second want to explore more types of Turducken-style code generation tasks to promote the practical application of our study, such as generating programs with regular expressions embedded in Java or Python and JavaScript embedded in HTML.

\begin{acknowledgements}
This work is supported by the National Natural Science Foundation of China (No.\ 61972197), the Natural Science Foundation of Jiangsu Province (No.\ BK20201292), the Collaborative Innovation Center of Novel Software Technology and Industrialization, the Open Project of Key Laboratory of Safety-Critical Software for Nanjing University of Aeronautics and Astronautics, Ministry of Industry and Information Technology (No.\ NJ2020022), and the Postgraduate Research \& Practice Innovation Program of Jiangsu Province (No.\ KYCX23\_0396).
T. Chen is partially supported by an oversea grant from the State Key Laboratory of Novel Software Technology, Nanjing University (KFKT2022A03), Birkbeck BEI School Project (EFFECT), and National Natural Science Foundation of China (No.\ 62272397).  
\end{acknowledgements}

%
\section*{Conflict of Interest}
The authors declare that they have no conflict of interest.

\section*{Data Availability Statements}
The datasets generated during and analyzed during the current study are available in the Github repository (\url{https://github.com/NTDXYG/TurduckenGen}).


\bibliographystyle{spbasic}
\bibliography{main}

\begin{thebibliography}{83}
\providecommand{\natexlab}[1]{#1}
\providecommand{\url}[1]{{#1}}
\providecommand{\urlprefix}{URL }
\expandafter\ifx\csname urlstyle\endcsname\relax
  \providecommand{\doi}[1]{DOI~\discretionary{}{}{}#1}\else
  \providecommand{\doi}{DOI~\discretionary{}{}{}\begingroup
  \urlstyle{rm}\Url}\fi
\providecommand{\eprint}[2][]{\url{#2}}

\bibitem[{Ahmad et~al(2021)Ahmad, Chakraborty, Ray, and
  Chang}]{ahmad2021unified}
Ahmad W, Chakraborty S, Ray B, Chang KW (2021) Unified pre-training for program
  understanding and generation. In: Proceedings of the 2021 Conference of the
  North American Chapter of the Association for Computational Linguistics:
  Human Language Technologies, pp 2655--2668

\bibitem[{Allamanis and Sutton(2013)}]{allamanis2013and}
Allamanis M, Sutton C (2013) Why, when, and what: analyzing stack overflow
  questions by topic, type, and code. In: 2013 10th Working conference on
  mining software repositories (MSR), IEEE, pp 53--56

\bibitem[{Bailey(2009)}]{bailey2009workshop}
Bailey MW (2009) Workshop on declarative aspects of multicore programming (damp
  2009) damp 2009

\bibitem[{Bogin et~al(2019)Bogin, Berant, and Gardner}]{bogin2019representing}
Bogin B, Berant J, Gardner M (2019) Representing schema structure with graph
  neural networks for text-to-sql parsing. In: Proceedings of the 57th Annual
  Meeting of the Association for Computational Linguistics, pp 4560--4565

\bibitem[{Chakraborty et~al(2022)Chakraborty, Ahmed, Ding, Devanbu, and
  Ray}]{chakraborty2022natgen}
Chakraborty S, Ahmed T, Ding Y, Devanbu PT, Ray B (2022) Natgen: generative
  pre-training by “naturalizing” source code. In: Proceedings of the 30th
  ACM Joint European Software Engineering Conference and Symposium on the
  Foundations of Software Engineering, pp 18--30

\bibitem[{Dahl et~al(1994)Dahl, Bates, Brown, Fisher, Hunicke-Smith, Pallett,
  Pao, Rudnicky, and Shriberg}]{dahl1994expanding}
Dahl DA, Bates M, Brown MK, Fisher WM, Hunicke-Smith K, Pallett DS, Pao C,
  Rudnicky A, Shriberg E (1994) Expanding the scope of the atis task: The
  atis-3 corpus. In: Human Language Technology: Proceedings of a Workshop held
  at Plainsboro, New Jersey, March 8-11, 1994

\bibitem[{Dauphin et~al(2017)Dauphin, Fan, Auli, and
  Grangier}]{dauphin2017language}
Dauphin YN, Fan A, Auli M, Grangier D (2017) Language modeling with gated
  convolutional networks. In: International conference on machine learning,
  PMLR, pp 933--941

\bibitem[{Dong et~al(2022)Dong, Jiang, Liu, Li, and
  Jin}]{https://doi.org/10.48550/arxiv.2211.00818}
Dong Y, Jiang X, Liu Y, Li G, Jin Z (2022) Codepad: Sequence-based code
  generation with pushdown automaton. \doi{10.48550/ARXIV.2211.00818},
  \urlprefix\url{https://arxiv.org/abs/2211.00818}

\bibitem[{Eghbali and Pradel(2022)}]{eghbali2022crystalbleu}
Eghbali A, Pradel M (2022) Crystalbleu: precisely and efficiently measuring the
  similarity of code. In: Proceedings of the ACM/IEEE 44th International
  Conference on Software Engineering: Companion Proceedings, pp 341--342

\bibitem[{Feng et~al(2020)Feng, Guo, Tang, Duan, Feng, Gong, Shou, Qin, Liu,
  Jiang et~al}]{feng2020codebert}
Feng Z, Guo D, Tang D, Duan N, Feng X, Gong M, Shou L, Qin B, Liu T, Jiang D,
  et~al (2020) Codebert: A pre-trained model for programming and natural
  languages. In: Findings of the Association for Computational Linguistics:
  EMNLP 2020, pp 1536--1547

\bibitem[{Fernandes and Bernardino(2015)}]{fernandes2015bigquery}
Fernandes S, Bernardino J (2015) What is bigquery? In: Proceedings of the 19th
  International Database Engineering \& Applications Symposium, pp 202--203

\bibitem[{Gao et~al(2021)Gao, Fisch, and Chen}]{gao2021making}
Gao T, Fisch A, Chen D (2021) Making pre-trained language models better
  few-shot learners. In: Joint Conference of the 59th Annual Meeting of the
  Association for Computational Linguistics and the 11th International Joint
  Conference on Natural Language Processing, ACL-IJCNLP 2021, Association for
  Computational Linguistics (ACL), pp 3816--3830

\bibitem[{Gifford and Lucassen(1986)}]{gifford1986integrating}
Gifford DK, Lucassen JM (1986) Integrating functional and imperative
  programming. In: Proceedings of the 1986 ACM Conference on LISP and
  Functional Programming, pp 28--38

\bibitem[{Gu et~al(2022)Gu, Han, Liu, and Huang}]{gu2022ppt}
Gu Y, Han X, Liu Z, Huang M (2022) Ppt: Pre-trained prompt tuning for few-shot
  learning. In: Proceedings of the 60th Annual Meeting of the Association for
  Computational Linguistics (Volume 1: Long Papers), pp 8410--8423

\bibitem[{Guo et~al(2021)Guo, Ren, Lu, Feng, Tang, Liu, Zhou, Duan,
  Svyatkovskiy, Fu et~al}]{guo2021graphcodebert}
Guo D, Ren S, Lu S, Feng Z, Tang D, Liu S, Zhou L, Duan N, Svyatkovskiy A, Fu
  S, et~al (2021) Graphcodebert: Pre-training code representations with data
  flow. In: ICLR

\bibitem[{Guo et~al(2022)Guo, Lu, Duan, Wang, Zhou, and Yin}]{guo2022unixcoder}
Guo D, Lu S, Duan N, Wang Y, Zhou M, Yin J (2022) Unixcoder: Unified
  cross-modal pre-training for code representation. In: Proceedings of the 60th
  Annual Meeting of the Association for Computational Linguistics (Volume 1:
  Long Papers), pp 7212--7225

\bibitem[{Hayati et~al(2018)Hayati, Olivier, Avvaru, Yin, Tomasic, and
  Neubig}]{hayati2018retrieval}
Hayati SA, Olivier R, Avvaru P, Yin P, Tomasic A, Neubig G (2018)
  Retrieval-based neural code generation. In: Proceedings of the 2018
  Conference on Empirical Methods in Natural Language Processing, pp 925--930

\bibitem[{Hu et~al(2020)Hu, Li, Xia, Lo, and Jin}]{hu2020deep}
Hu X, Li G, Xia X, Lo D, Jin Z (2020) Deep code comment generation with hybrid
  lexical and syntactical information. Empirical Software Engineering
  25(3):2179--2217

\bibitem[{Hu et~al(2021)Hu, Gao, Xia, Lo, and Yang}]{9678552}
Hu X, Gao Z, Xia X, Lo D, Yang X (2021) Automating user notice generation for
  smart contract functions. In: 2021 36th IEEE/ACM International Conference on
  Automated Software Engineering (ASE), pp 5--17,
  \doi{10.1109/ASE51524.2021.9678552}

\bibitem[{Hu et~al(2022)Hu, Xia, Lo, Wan, Chen, and
  Zimmermann}]{DBLP:conf/icse/HuX0WCZ22}
Hu X, Xia X, Lo D, Wan Z, Chen Q, Zimmermann T (2022) Practitioners'
  expectations on automated code comment generation. In: 44th {IEEE/ACM} 44th
  International Conference on Software Engineering, {ICSE} 2022, Pittsburgh,
  PA, USA, May 25-27, 2022, {ACM}, pp 1693--1705, \doi{10.1145/3510003.3510152}

\bibitem[{Huang et~al(2021)Huang, Wang, Wang, Dong, and
  Xiao}]{huang2021relation}
Huang J, Wang Y, Wang Y, Dong Y, Xiao Y (2021) Relation aware
  semi-autoregressive semantic parsing for nl2sql. arXiv preprint
  arXiv:210800804

\bibitem[{Huang et~al(2022)Huang, Wang, Zhang, Yan, Cui, Inala, Clement, Duan,
  and Gao}]{huang2022execution}
Huang J, Wang C, Zhang J, Yan C, Cui H, Inala JP, Clement C, Duan N, Gao J
  (2022) Execution-based evaluation for data science code generation models.
  arXiv preprint arXiv:221109374

\bibitem[{Husain et~al(2019)Husain, Wu, Gazit, Allamanis, and
  Brockschmidt}]{husain2019codesearchnet}
Husain H, Wu HH, Gazit T, Allamanis M, Brockschmidt M (2019) Codesearchnet
  challenge: Evaluating the state of semantic code search. arXiv preprint
  arXiv:190909436

\bibitem[{Hussain et~al(2020{\natexlab{a}})Hussain, Huang, Zhou, and
  Wang}]{hussain2020codegru}
Hussain Y, Huang Z, Zhou Y, Wang S (2020{\natexlab{a}}) Codegru: Context-aware
  deep learning with gated recurrent unit for source code modeling. Information
  and Software Technology 125:106309

\bibitem[{Hussain et~al(2020{\natexlab{b}})Hussain, Huang, Zhou, and
  Wang}]{hussain2020deep}
Hussain Y, Huang Z, Zhou Y, Wang S (2020{\natexlab{b}}) Deep transfer learning
  for source code modeling. International Journal of Software Engineering and
  Knowledge Engineering 30(05):649--668

\bibitem[{Hussain et~al(2021)Hussain, Huang, and Zhou}]{hussain2021improving}
Hussain Y, Huang Z, Zhou Y (2021) Improving source code suggestion with code
  embedding and enhanced convolutional long short-term memory. IET Software
  15(3):199--213

\bibitem[{Iyer et~al(2017)Iyer, Konstas, Cheung, Krishnamurthy, and
  Zettlemoyer}]{iyer2017learning}
Iyer S, Konstas I, Cheung A, Krishnamurthy J, Zettlemoyer L (2017) Learning a
  neural semantic parser from user feedback. In: Proceedings of the 55th Annual
  Meeting of the Association for Computational Linguistics (Volume 1: Long
  Papers), pp 963--973

\bibitem[{Klein et~al(2017)Klein, Kim, Deng, Senellart, and
  Rush}]{klein2017opennmt}
Klein G, Kim Y, Deng Y, Senellart J, Rush AM (2017) Opennmt: Open-source
  toolkit for neural machine translation. In: Proceedings of ACL 2017, System
  Demonstrations, pp 67--72

\bibitem[{Le et~al(2022)Le, Wang, Gotmare, Savarese, and Hoi}]{le2022coderl}
Le H, Wang Y, Gotmare AD, Savarese S, Hoi SC (2022) Coderl: Mastering code
  generation through pretrained models and deep reinforcement learning. arXiv
  preprint arXiv:220701780

\bibitem[{Legendre(2005)}]{legendre2005species}
Legendre P (2005) Species associations: the kendall coefficient of concordance
  revisited. Journal of agricultural, biological, and environmental statistics
  10(2):226--245

\bibitem[{Li and Liang(2021)}]{li2021prefix}
Li XL, Liang P (2021) Prefix-tuning: Optimizing continuous prompts for
  generation. In: Proceedings of the 59th Annual Meeting of the Association for
  Computational Linguistics and the 11th International Joint Conference on
  Natural Language Processing (Volume 1: Long Papers), pp 4582--4597

\bibitem[{Liang et~al(2021)Liang, Sun, Zhu, Zhang, Yu, Xiong, and
  Zhang}]{liang2021lyra}
Liang Q, Sun Z, Zhu Q, Zhang W, Yu L, Xiong Y, Zhang L (2021) Lyra: A benchmark
  for turducken-style code generation. arXiv preprint arXiv:210812144

\bibitem[{Liguori et~al(2021)Liguori, Al-Hossami, Orbinato, Natella, Shaikh,
  Cotroneo, and Cukic}]{liguori2021evil}
Liguori P, Al-Hossami E, Orbinato V, Natella R, Shaikh S, Cotroneo D, Cukic B
  (2021) Evil: exploiting software via natural language. In: 2021 IEEE 32nd
  International Symposium on Software Reliability Engineering (ISSRE), IEEE, pp
  321--332

\bibitem[{Lin et~al(2020)Lin, Socher, and Xiong}]{lin2020bridging}
Lin XV, Socher R, Xiong C (2020) Bridging textual and tabular data for
  cross-domain text-to-sql semantic parsing. In: Findings of the Association
  for Computational Linguistics: EMNLP 2020, pp 4870--4888

\bibitem[{Ling et~al(2016)Ling, Blunsom, Grefenstette, Hermann,
  Ko{\v{c}}isk{\`y}, Wang, and Senior}]{ling2016latent}
Ling W, Blunsom P, Grefenstette E, Hermann KM, Ko{\v{c}}isk{\`y} T, Wang F,
  Senior A (2016) Latent predictor networks for code generation. In:
  Proceedings of the 54th Annual Meeting of the Association for Computational
  Linguistics (Volume 1: Long Papers), pp 599--609

\bibitem[{Liu et~al(2020{\natexlab{a}})Liu, Li, Zhao, and Jin}]{liu2020multi}
Liu F, Li G, Zhao Y, Jin Z (2020{\natexlab{a}}) Multi-task learning based
  pre-trained language model for code completion. In: Proceedings of the 35th
  IEEE/ACM International Conference on Automated Software Engineering, pp
  473--485

\bibitem[{Liu et~al(2022)Liu, Li, Wei, Xia, Fu, and Jin}]{liu2022unified}
Liu F, Li G, Wei B, Xia X, Fu Z, Jin Z (2022) A unified multi-task learning
  model for ast-level and token-level code completion. Empirical Software
  Engineering 27(4):1--38

\bibitem[{Liu et~al(2023{\natexlab{a}})Liu, Li, and Zhang}]{liu2023syntax}
Liu F, Li J, Zhang L (2023{\natexlab{a}}) Syntax and domain aware model for
  unsupervised program translation. arXiv preprint arXiv:230203908

\bibitem[{Liu et~al(2023{\natexlab{b}})Liu, Yuan, Fu, Jiang, Hayashi, and
  Neubig}]{liu2023pre}
Liu P, Yuan W, Fu J, Jiang Z, Hayashi H, Neubig G (2023{\natexlab{b}})
  Pre-train, prompt, and predict: A systematic survey of prompting methods in
  natural language processing. ACM Computing Surveys 55(9):1--35

\bibitem[{Liu et~al(2020{\natexlab{b}})Liu, Chen, Chen, Lou, Chen, Zhou, and
  Zhang}]{liu2020you}
Liu Q, Chen Y, Chen B, Lou JG, Chen Z, Zhou B, Zhang D (2020{\natexlab{b}}) You
  impress me: Dialogue generation via mutual persona perception. In:
  Proceedings of the 58th Annual Meeting of the Association for Computational
  Linguistics, pp 1417--1427

\bibitem[{Liu et~al(2023{\natexlab{c}})Liu, Tantithamthavorn, Liu, and
  Li}]{liu2023reliability}
Liu Y, Tantithamthavorn C, Liu Y, Li L (2023{\natexlab{c}}) On the reliability
  and explainability of automated code generation approaches. arXiv preprint
  arXiv:230209587

\bibitem[{Lloyd(1994)}]{lloyd1994practical}
Lloyd JW (1994) Practical advtanages of declarative programming. In: GULP-PRODE
  (1), pp 18--30

\bibitem[{Longpre et~al(2023)Longpre, Hou, Vu, Webson, Chung, Tay, Zhou, Le,
  Zoph, Wei et~al}]{longpre2023flan}
Longpre S, Hou L, Vu T, Webson A, Chung HW, Tay Y, Zhou D, Le QV, Zoph B, Wei
  J, et~al (2023) The flan collection: Designing data and methods for effective
  instruction tuning. arXiv preprint arXiv:230113688

\bibitem[{Lu et~al(2021{\natexlab{a}})Lu, Guo, Ren, Huang, Svyatkovskiy,
  Blanco, Clement, Drain, Jiang, Tang et~al}]{lu2021codexglue}
Lu S, Guo D, Ren S, Huang J, Svyatkovskiy A, Blanco A, Clement C, Drain D,
  Jiang D, Tang D, et~al (2021{\natexlab{a}}) Codexglue: A machine learning
  benchmark dataset for code understanding and generation. arXiv preprint
  arXiv:210204664

\bibitem[{Lu et~al(2021{\natexlab{b}})Lu, Guo, Ren, Huang, Svyatkovskiy,
  Blanco, Clement, Drain, Jiang, Tang, Li, Zhou, Shou, Zhou, Tufano, Gong,
  Zhou, Duan, Sundaresan, Deng, Fu, and Liu}]{DBLP:conf/nips/LuGRHSBCDJTLZSZ21}
Lu S, Guo D, Ren S, Huang J, Svyatkovskiy A, Blanco A, Clement CB, Drain D,
  Jiang D, Tang D, Li G, Zhou L, Shou L, Zhou L, Tufano M, Gong M, Zhou M, Duan
  N, Sundaresan N, Deng SK, Fu S, Liu S (2021{\natexlab{b}}) Codexglue: {A}
  machine learning benchmark dataset for code understanding and generation. In:
  Vanschoren J, Yeung S (eds) Proceedings of the Neural Information Processing
  Systems Track on Datasets and Benchmarks 1, NeurIPS Datasets and Benchmarks
  2021, December 2021, virtual

\bibitem[{Mahmud et~al(2015)Mahmud, Hasan, Ahmed, and Chak}]{mahmud2015rule}
Mahmud T, Hasan KA, Ahmed M, Chak THC (2015) A rule based approach for nlp
  based query processing. In: 2015 2nd International Conference on Electrical
  Information and Communication Technologies (EICT), IEEE, pp 78--82

\bibitem[{Mou et~al(2015)Mou, Men, Li, Zhang, and Jin}]{mou2015end}
Mou L, Men R, Li G, Zhang L, Jin Z (2015) On end-to-end program generation from
  user intention by deep neural networks. arXiv preprint arXiv:151007211

\bibitem[{Niu et~al(2022)Niu, Li, Ng, Ge, Huang, and Luo}]{niu2022spt}
Niu C, Li C, Ng V, Ge J, Huang L, Luo B (2022) Spt-code: sequence-to-sequence
  pre-training for learning source code representations. In: Proceedings of the
  44th International Conference on Software Engineering, pp 2006--2018

\bibitem[{Papineni et~al(2002)Papineni, Roukos, Ward, and
  Zhu}]{papineni2002bleu}
Papineni K, Roukos S, Ward T, Zhu WJ (2002) Bleu: a method for automatic
  evaluation of machine translation. In: Proceedings of the 40th annual meeting
  of the Association for Computational Linguistics, pp 311--318

\bibitem[{Popescu et~al(2003)Popescu, Etzioni, and Kautz}]{popescu2003towards}
Popescu AM, Etzioni O, Kautz H (2003) Towards a theory of natural language
  interfaces to databases. In: Proceedings of the 8th international conference
  on Intelligent user interfaces, pp 149--157

\bibitem[{Radford et~al(2019)Radford, Wu, Child, Luan, Amodei, Sutskever
  et~al}]{radford2019language}
Radford A, Wu J, Child R, Luan D, Amodei D, Sutskever I, et~al (2019) Language
  models are unsupervised multitask learners. OpenAI blog 1(8):9

\bibitem[{Raffel et~al(2020)Raffel, Shazeer, Roberts, Lee, Narang, Matena,
  Zhou, Li, and Liu}]{raffel2020exploring}
Raffel C, Shazeer N, Roberts A, Lee K, Narang S, Matena M, Zhou Y, Li W, Liu PJ
  (2020) Exploring the limits of transfer learning with a unified text-to-text
  transformer. The Journal of Machine Learning Research 21(1):5485--5551

\bibitem[{Ren et~al(2020)Ren, Guo, Lu, Zhou, Liu, Tang, Sundaresan, Zhou,
  Blanco, and Ma}]{ren2020codebleu}
Ren S, Guo D, Lu S, Zhou L, Liu S, Tang D, Sundaresan N, Zhou M, Blanco A, Ma S
  (2020) Codebleu: a method for automatic evaluation of code synthesis. arXiv
  preprint arXiv:200910297

\bibitem[{Rubin and Berant(2021)}]{rubin2021smbop}
Rubin O, Berant J (2021) Smbop: Semi-autoregressive bottom-up semantic parsing.
  In: Proceedings of the 5th Workshop on Structured Prediction for NLP (SPNLP
  2021), pp 12--21

\bibitem[{S{\'a}nchez-Cartagena et~al(2021)S{\'a}nchez-Cartagena,
  Espl{\`a}-Gomis, P{\'e}rez-Ortiz, and
  S{\'a}nchez-Mart{\'\i}nez}]{sanchez2021rethinking}
S{\'a}nchez-Cartagena VM, Espl{\`a}-Gomis M, P{\'e}rez-Ortiz JA,
  S{\'a}nchez-Mart{\'\i}nez F (2021) Rethinking data augmentation for
  low-resource neural machine translation: A multi-task learning approach. In:
  Proceedings of the 2021 Conference on Empirical Methods in Natural Language
  Processing, pp 8502--8516

\bibitem[{Scholak et~al(2021)Scholak, Schucher, and
  Bahdanau}]{scholak2021picard}
Scholak T, Schucher N, Bahdanau D (2021) Picard: Parsing incrementally for
  constrained auto-regressive decoding from language models. In: Proceedings of
  the 2021 Conference on Empirical Methods in Natural Language Processing, pp
  9895--9901

\bibitem[{Sun et~al(2019)Sun, Zhu, Mou, Xiong, Li, and Zhang}]{sun2019grammar}
Sun Z, Zhu Q, Mou L, Xiong Y, Li G, Zhang L (2019) A grammar-based structural
  cnn decoder for code generation. In: Proceedings of the AAAI conference on
  artificial intelligence, vol~33, pp 7055--7062

\bibitem[{Sun et~al(2020)Sun, Zhu, Xiong, Sun, Mou, and Zhang}]{sun2020treegen}
Sun Z, Zhu Q, Xiong Y, Sun Y, Mou L, Zhang L (2020) Treegen: A tree-based
  transformer architecture for code generation. In: Proceedings of the AAAI
  Conference on Artificial Intelligence, vol~34, pp 8984--8991

\bibitem[{Vaswani et~al(2017)Vaswani, Shazeer, Parmar, Uszkoreit, Jones, Gomez,
  Kaiser, and Polosukhin}]{vaswani2017attention}
Vaswani A, Shazeer N, Parmar N, Uszkoreit J, Jones L, Gomez AN, Kaiser {\L},
  Polosukhin I (2017) Attention is all you need. Advances in neural information
  processing systems 30

\bibitem[{Wang et~al(2020)Wang, Shin, Liu, Polozov, and
  Richardson}]{wang2020rat}
Wang B, Shin R, Liu X, Polozov O, Richardson M (2020) Rat-sql: Relation-aware
  schema encoding and linking for text-to-sql parsers. In: Proceedings of the
  58th Annual Meeting of the Association for Computational Linguistics, pp
  7567--7578

\bibitem[{Wang et~al(2022{\natexlab{a}})Wang, Yang, Gao, Peng, Zhang, and
  Lyu}]{wang2022no}
Wang C, Yang Y, Gao C, Peng Y, Zhang H, Lyu MR (2022{\natexlab{a}}) No more
  fine-tuning? an experimental evaluation of prompt tuning in code
  intelligence. In: Proceedings of the 30th ACM Joint European Software
  Engineering Conference and Symposium on the Foundations of Software
  Engineering, pp 382--394

\bibitem[{Wang et~al(2021{\natexlab{a}})Wang, Yu, Li, Dong, Wang, and
  Qing}]{wang2021mulcode}
Wang D, Yu Y, Li S, Dong W, Wang J, Qing L (2021{\natexlab{a}}) Mulcode: A
  multi-task learning approach for source code understanding. In: 2021 IEEE
  International Conference on Software Analysis, Evolution and Reengineering
  (SANER), IEEE, pp 48--59

\bibitem[{Wang et~al(2022{\natexlab{b}})Wang, Wang, Wan, Mi, Li, Zhou, Liu, Wu,
  Jiang, and Liu}]{wang2022compilable}
Wang X, Wang Y, Wan Y, Mi F, Li Y, Zhou P, Liu J, Wu H, Jiang X, Liu Q
  (2022{\natexlab{b}}) Compilable neural code generation with compiler
  feedback. In: Findings of the Association for Computational Linguistics: ACL
  2022, pp 9--19

\bibitem[{Wang et~al(2021{\natexlab{b}})Wang, Wang, Joty, and
  Hoi}]{wang2021codet5}
Wang Y, Wang W, Joty S, Hoi SC (2021{\natexlab{b}}) Codet5: Identifier-aware
  unified pre-trained encoder-decoder models for code understanding and
  generation. In: Proceedings of the 2021 Conference on Empirical Methods in
  Natural Language Processing, pp 8696--8708

\bibitem[{Wei et~al(2019)Wei, Li, Xia, Fu, and Jin}]{wei2019code}
Wei B, Li G, Xia X, Fu Z, Jin Z (2019) Code generation as a dual task of code
  summarization. Advances in neural information processing systems 32

\bibitem[{Wilcoxon(1992)}]{wilcoxon1992individual}
Wilcoxon F (1992) Individual comparisons by ranking methods. In: Breakthroughs
  in statistics, Springer, pp 196--202

\bibitem[{Wiseman and Rush(2016)}]{wiseman2016sequence}
Wiseman S, Rush AM (2016) Sequence-to-sequence learning as beam-search
  optimization. In: Proceedings of the 2016 Conference on Empirical Methods in
  Natural Language Processing, pp 1296--1306

\bibitem[{Xie et~al(2021)Xie, Ye, Sun, and Zhang}]{xie2021exploiting}
Xie R, Ye W, Sun J, Zhang S (2021) Exploiting method names to improve code
  summarization: A deliberation multi-task learning approach. In: 2021 IEEE/ACM
  29th International Conference on Program Comprehension (ICPC), IEEE, pp
  138--148

\bibitem[{Xu et~al(2022)Xu, Vasilescu, and Neubig}]{xu2022ide}
Xu FF, Vasilescu B, Neubig G (2022) In-ide code generation from natural
  language: Promise and challenges. ACM Transactions on Software Engineering
  and Methodology (TOSEM) 31(2):1--47

\bibitem[{Xuan et~al(2021)Xuan, Wang, Wang, Wen, and Dong}]{xuan2021sead}
Xuan K, Wang Y, Wang Y, Wen Z, Dong Y (2021) Sead: End-to-end text-to-sql
  generation with schema-aware denoising. arXiv preprint arXiv:210507911

\bibitem[{Yang et~al(2021{\natexlab{a}})Yang, Chen, Cao, Xu, Cui, Yu, and
  Liu}]{yang2021comformer}
Yang G, Chen X, Cao J, Xu S, Cui Z, Yu C, Liu K (2021{\natexlab{a}}) Comformer:
  Code comment generation via transformer and fusion method-based hybrid code
  representation. In: 2021 8th International Conference on Dependable Systems
  and Their Applications (DSA), IEEE, pp 30--41

\bibitem[{Yang et~al(2021{\natexlab{b}})Yang, Zhou, Chen, and
  Yu}]{yang2021fine}
Yang G, Zhou Y, Chen X, Yu C (2021{\natexlab{b}}) Fine-grained pseudo-code
  generation method via code feature extraction and transformer. In: 2021 28th
  Asia-Pacific Software Engineering Conference (APSEC), IEEE, pp 213--222

\bibitem[{Yang et~al(2022{\natexlab{a}})Yang, Chen, Zhou, and
  Yu}]{yang2022dualsc}
Yang G, Chen X, Zhou Y, Yu C (2022{\natexlab{a}}) Dualsc: Automatic generation
  and summarization of shellcode via transformer and dual learning. arXiv
  preprint arXiv:220209785

\bibitem[{Yang et~al(2022{\natexlab{b}})Yang, Chen, Zhou, and
  Yu}]{DBLP:conf/wcre/YangCZY22}
Yang G, Chen X, Zhou Y, Yu C (2022{\natexlab{b}}) Dualsc: Automatic generation
  and summarization of shellcode via transformer and dual learning. In: {IEEE}
  International Conference on Software Analysis, Evolution and Reengineering,
  {SANER} 2022, Honolulu, HI, USA, March 15-18, 2022, {IEEE}, pp 361--372,
  \doi{10.1109/SANER53432.2022.00052}

\bibitem[{Yang et~al(2022{\natexlab{c}})Yang, Zhou, Chen, Zhang, Han, and
  Chen}]{yang2022exploitgen}
Yang G, Zhou Y, Chen X, Zhang X, Han T, Chen T (2022{\natexlab{c}}) Exploitgen:
  Template-augmented exploit code generation based on codebert. Journal of
  Systems and Software p 111577

\bibitem[{Yang et~al(2023)Yang, Zhou, Chen, Zhang, Han, and
  Chen}]{yang2023exploitgen}
Yang G, Zhou Y, Chen X, Zhang X, Han T, Chen T (2023) Exploitgen:
  Template-augmented exploit code generation based on codebert. Journal of
  Systems and Software 197:111577

\bibitem[{Yin and Neubig(2018)}]{yin2018tranx}
Yin P, Neubig G (2018) Tranx: A transition-based neural abstract syntax parser
  for semantic parsing and code generation. In: Proceedings of the 2018
  Conference on Empirical Methods in Natural Language Processing: System
  Demonstrations, pp 7--12

\bibitem[{Yu et~al(2018{\natexlab{a}})Yu, Li, Zhang, Zhang, and
  Radev}]{yu2018typesql}
Yu T, Li Z, Zhang Z, Zhang R, Radev D (2018{\natexlab{a}}) Typesql:
  Knowledge-based type-aware neural text-to-sql generation. In: Proceedings of
  the 2018 Conference of the North American Chapter of the Association for
  Computational Linguistics: Human Language Technologies, Volume 2 (Short
  Papers), pp 588--594

\bibitem[{Yu et~al(2018{\natexlab{b}})Yu, Zhang, Yang, Yasunaga, Wang, Li, Ma,
  Li, Yao, Roman et~al}]{yu2018spider}
Yu T, Zhang R, Yang K, Yasunaga M, Wang D, Li Z, Ma J, Li I, Yao Q, Roman S,
  et~al (2018{\natexlab{b}}) Spider: A large-scale human-labeled dataset for
  complex and cross-domain semantic parsing and text-to-sql task. In:
  Proceedings of the 2018 Conference on Empirical Methods in Natural Language
  Processing, pp 3911--3921

\bibitem[{Yu et~al(2019{\natexlab{a}})Yu, Zhang, Er, Li, Xue, Pang, Lin, Tan,
  Shi, Li et~al}]{yu2019cosql}
Yu T, Zhang R, Er H, Li S, Xue E, Pang B, Lin XV, Tan YC, Shi T, Li Z, et~al
  (2019{\natexlab{a}}) Cosql: A conversational text-to-sql challenge towards
  cross-domain natural language interfaces to databases. In: Proceedings of the
  2019 Conference on Empirical Methods in Natural Language Processing and the
  9th International Joint Conference on Natural Language Processing
  (EMNLP-IJCNLP), pp 1962--1979

\bibitem[{Yu et~al(2019{\natexlab{b}})Yu, Zhang, Yasunaga, Tan, Lin, Li, Er,
  Li, Pang, Chen et~al}]{yu2019sparc}
Yu T, Zhang R, Yasunaga M, Tan YC, Lin XV, Li S, Er H, Li I, Pang B, Chen T,
  et~al (2019{\natexlab{b}}) Sparc: Cross-domain semantic parsing in context.
  In: Proceedings of the 57th Annual Meeting of the Association for
  Computational Linguistics, pp 4511--4523

\bibitem[{Zelle and Mooney(1996)}]{zelle1996learning}
Zelle JM, Mooney RJ (1996) Learning to parse database queries using inductive
  logic programming. In: Proceedings of the national conference on artificial
  intelligence, pp 1050--1055

\bibitem[{Zhong et~al(2017)Zhong, Xiong, and Socher}]{zhong2017seq2sql}
Zhong V, Xiong C, Socher R (2017) Seq2sql: Generating structured queries from
  natural language using reinforcement learning. arXiv preprint arXiv:170900103

\end{thebibliography}

\end{document}